\documentclass[useAMS]{mn2e}
\usepackage{epsfig}
\usepackage{epsf}
\usepackage{graphicx}

\title[SN~1987A-like Ring Around SBW1]{The Ring Nebula Around the Blue
  Supergiant SBW1: Pre-Explosion Snapshot of a SN~1987A
  Twin\thanks{Based in part on observations made with the NASA/ESA
    {\it Hubble Space Telescope}, obtained at the Space Telescope
    Science Institute, which is operated by the Association of
    Universities for Research in Astronomy, Inc.\ (AURA), under NASA
    contract NAS5-26555.}\thanks{Based in part on observations
    obtained at the Gemin Observatory, which is operated by AURA,
    under a cooperative agreement with the National Science Foundation
    (NSF) on behalf of the Gemini partnership, which comprises the NSF
    (US), the Particle Physics and Astronomy Research Council (UK),
    the National Research Council (Canada), CONICYT (Chile), the
    Australian Research Council (Australia), CNPq (Brazil), and
    CONICET (Argentina).}}

\author[Smith et al.]{Nathan Smith$^1$\thanks{Email:
    nathans@as.arizona.edu}, W.\ David Arnett$^1$, John Bally$^2$,
  Adam Ginsburg$^2$ \& Alexei V. \newauthor Filippenko$^3$ \\
  $^1$Steward Observatory, 933 N. Cherry Ave., Tucson, AZ 85721, USA
  \\ $^2$Center for Astrophysics and Space Astronomy, Campus Box 389,
  University of Colorado, Boulder, CO 80309, USA \\ $^3$Department of
  Astronomy, University of California, Berkeley, CA 94720-3411, USA}

\begin{document}
\date{Accepted 2012 Nov 14, Received 2012 Oct 17, in original form 2012 Oct 17}
\pagerange{\pageref{firstpage}--\pageref{lastpage}} \pubyear{2002}
\def\arcdeg{\degr}
\maketitle
\label{firstpage}

\begin{abstract}

  SBW1 is a B-type supergiant surrounded by a ring nebula that is a
  nearby twin of SN~1987A's progenitor and its circumstellar ring.  We
  present images and spectra of SBW1 obtained with the {\it Hubble
    Space Telescope} ({\it HST}), the {\it Spitzer Space Telescope},
  and Gemini South.  {\it HST} images of SBW1 do not exhibit long
  Rayleigh-Taylor (R-T) fingers, which are presumed to cause the
  ``hotspots'' in the SN~1987A ring when impacted by the blast wave,
  but instead show a geometrically thin ($\Delta R/R \la 0.05$) clumpy
  ring.  The radial mass distribution and size scales of
  inhomogeneities in SBW1's ring closely resemble those in the
  SN~1987A ring, but the more complete disk expected to reside at the
  base of the R-T fingers is absent in SBW1.  This structure may
  explain why portions of the SN~1987A ring between the hotspots have
  not yet brightened, more than 15 years after the first hotspots
  appeared.  The model we suggest does not require a fast wind
  colliding with a previous red supergiant wind, because a slowly
  expanding equatorial ring may be ejected by a rotating blue
  supergiant star or in a close binary system.  More surprisingly,
  high-resolution images of SBW1 also reveal diffuse emission filling
  the interior of the ring seen in H$\alpha$ and in thermal-infrared
  (IR) emission; $\sim 190$\,K dust dominates the 8--20\,$\mu$m
  luminosity (but contains only $10^{-5}$\,M$_{\odot}$ of dust).
  Cooler ($\sim 85$\,K) dust resides in the equatorial ring itself
  (and has a dust mass of at least $5 \times 10^{-3}$\,M$_{\odot}$).
  Diffuse emission extends inward to $\sim$1{\arcsec} from the central
  star, where a paucity of H$\alpha$ and IR emission suggests an inner
  hole excavated by the B-supergiant wind.  We propose that diffuse
  emission inside the ring arises from an ionised flow of material
  photoevaporated from the dense ring, and its pressure prevents the
  B-supergiant wind from advancing in the equatorial plane.  This
  inner emission could correspond to a structure hypothesised to
  reside around Sk$-$69{\arcdeg}202 that was never directly detected.
  If this interpretation is correct, it would suggest that
  photoionisation can play an important dynamical role in shaping the
  ring nebula, and we speculate that this might help explain the
  origin of the polar rings around SN~1987A. In effect, the
  photoevaporative flow shields the outer bipolar nebula at low
  latitudes, whereas the blue-supergiant wind expands freely out the
  poles and clears away the polar caps of the nebula; the polar rings
  reside at the intersection of these two zones.

\end{abstract}


\section{INTRODUCTION}

SN 1987A in the Large Magellanic Cloud (LMC) is one of those vexing
examples of Murphy's law, where the nearest and best-studied supernova
(SN) in modern times also appears unique when compared to the
population of known extragalactic core-collapse supernovae (SNe).  It
is an oddball, consisting of a peculiar SN type having a slow rise to
peak, attributed to the relatively compact nature of its (also unusual
and unexpected) blue-supergiant (BSG) progenitor (Arnett 1987, 1989;
Arnett et al.\ 1989).  SN~1987A is the only SN near enough to obtain a
clear picture of its spatially resolved circumstellar nebula, but its
bizarre triple-ring nebula still defies adequate explanation (Luo \&
McCray 1991; Blondin \& Lundqvist 1993; Martin \& Arnett 1995; Collins
et al.\ 1999; Morris \& Podsiadlowski 2009), and does not conform to
the hourglass structures commonly seen in bipolar planetary nebulae
and bipolar nebulae around massive stars.

There are, however, a few objects known that appear similar to
SN~1987A in that they have prominent equatorial ring nebulae with
bipolar lobes or rings surrounding BSG central stars.  In this paper,
we present a detailed study of the recently discovered ring nebula
SBW1 (Smith et al.\ 2007).  Two other known bipolar ring nebulae
around massive stars are HD~168625 and Sher~25, which are discussed in
detail elsewhere (Smith 2007; Brandner et al.\ 1997).  Of these three,
SBW1 appears to most closely resemble the nebula around SN~1987A and
the physical parameters of its progenitor star.

Before massive stars explode as SNe, they can shed considerable mass
as they attempt to remove their outer H-rich envelope. If they fail to
shed their H envelopes, they will typically remain as red supergiants
(RSGs) and will die as SNe of Type II.  One expects single stars in
the lower initial mass ranges of core-collapse SNe
(8--20\,M$_{\odot}$) to retain their H envelopes and explode as RSGs,
and this expectation is largely confirmed by pre-explosion detections
of progenitor stars of SNe~II-P (e.g., Smartt 2009; Leonard 2011, and
references therein).  SN~1987A challenged our view of stellar
evolution because it was a SN~II from an explosion of a BSG, not a
RSG, with an initial mass of 18--20\,M$_{\odot}$ (see Arnett et al.\
1989, and references therein). The reason that the progenitor was a
BSG is still unclear, but various scenarios involving close binary
evolution, binary mergers, rapid rotation, and enhanced mass loss have
been suggested.

Pre-explosion data for SN1987A's progenitor star Sk$-$69$\arcdeg$202
establish that it appeared to be a fairly normal B3~I supergiant
(Walborn et al.\ 1989; Rousseau et al.\ 1978). Therefore, our
conjectures about the progenitor and its pre-SN mass loss depend
heavily on studies of the remarkable ring nebula surrounding the SN,
made famous in early {\it Hubble Space Telescope (HST)} images (e.g.,
Plait et al.\ 1995; Burrows et al.\ 1995). Kinematic studies of the
nebula's expansion indicate that it was ejected by the progenitor star
roughly $10^4$\,yr before exploding as a SN (Meaburn et al.\ 1995;
Crotts \& Heathcote 2000). The ejection and shaping mechanisms of this
nebula are intimately linked to the star's peculiar evolution just
before explosion, but our understanding of that process is still
tenuous.  The poor understanding of how the ring nebula formed has
become a more pressing problem in recent years. We are now lucky
enough to witness a spectacular collision as the blast wave of the SN
overtakes the ring nebula ejected by the progenitor, predicted shortly
after the discovery of the nebula (Luo \& McCray 1991).

The SN~1987A blast wave first began to collide with the dense
circumstellar ring in 1997, heralded by the appearance of new ``hot
spots'' in the ring (Sonneborn et al.\ 1998; Michael et al.\ 1998,
1999).  These hotspots had broader line widths than the rest of the
ring, confirming that they were bright because a shock was being
driven into the dense clumps.  These clumps were thought to be the
ends of long ``fingers'' created by Rayleigh-Taylor (R-T)
instabilities at the contact discontinuity between the slow RSG wind
and the fast BSG wind.  Since the hotspots first appeared in 1997,
many more spots have brightened all around the ring (Sugerman et al.\
2002), although the hotspots have not yet merged into a contiguous
bright ring as one might expect when the blast wave catches up to gas
in between the R-T fingers.

In the decade between explosion and the start of this collision with
the ring, the blast wave was expanding through the relatively
low-density region interior to the ring.  Radio emission and
hydrodynamic models suggest that the BSG wind had a surprisingly low
mass-loss rate of $\sim 10^{-7}$\,M$_{\odot}$\,yr$^{-1}$
(Stavely-Smith et al.\ 1993; Blondin \& Lundqvist 1993; Martin \&
Arnett 1995; Chevalier \& Dwarkadas 1995). The expansion rate of the
blast wave through this low-density progenitor wind was fast at first,
but then slowed (Gaensler et al.\ 2000), attributed to the shock
running into a higher-density H~{\sc ii} region caused by photoionised
material from the dense RSG wind (Chevalier \& Dwarkadas 1995; Meyer
1997).  While the existence of this H~{\sc ii} region can account for
some observed characteristics of the blast-wave expansion, emission
from this feature itself was not directly observed before it was hit
by the blast wave.  One of the key results from our analysis below is
that such a feature is seen directly in the similar nebula around
SBW1.

\begin{figure*}\begin{center}
    \includegraphics[width=6.4in]{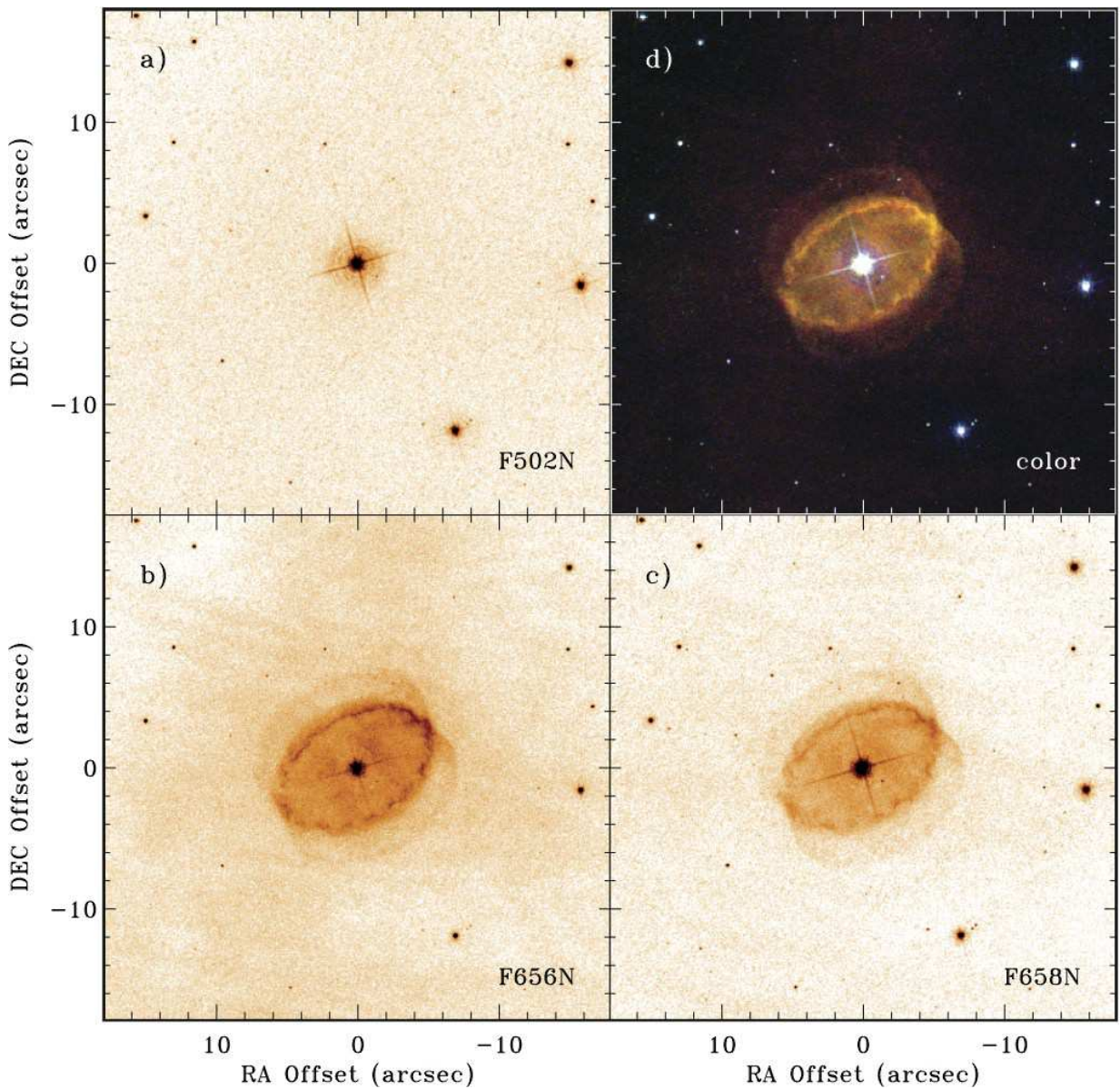}
\end{center}
\caption{The new {\it HST}/WFC3 images of SBW1.  Panels (a), (b), and
  (c) are taken in the F502N [O~{\sc iii}] $\lambda$5007,
  F656N H$\alpha$, and F658N [N~{\sc ii}] $\lambda$6584 filters,
  respectively, displayed in false colour.  Panel (d) is a colour
  composite of the three {\it HST}/WFC3 images, with F502N in blue,
  F658N in green, and F656N in red.  The origin is at the position of
  the star, at $\alpha_{\rm J2000}$ = 10$^h$40$^m$18$\fs$60,
  $\delta_{\rm J2000}$ = $-$59$\arcdeg$49$\arcmin$12$\farcs$5. }
\label{fig:hst}
\end{figure*}

\begin{table}\begin{center}\begin{minipage}{3.3in}
      \caption{New Observations of SBW1}
\scriptsize
\begin{tabular}{@{}llcl}\hline\hline
UT Date        &Tel./Instr.      &Filter  &Exp.  \\ \hline
2009 Dec. 08 &{\it HST}/WFC3  &F502N    &2$\times$210 s \\
2009 Dec. 08 &{\it HST}/WFC3  &F656N    &4$\times$210 s \\
2009 Dec. 08 &{\it HST}/WFC3  &F658N    &4$\times$240 s \\
2010 May 18  &{\it HST}/STIS  &G750M/6581      &2$\times$2070 s  \\
2010 Mar. 29 &Gemini South/T-ReCS &8.8 $\mu$m  &4$\times$1570 s   \\
2008 Mar. 30 &Gemini South/T-ReCS &11.7 $\mu$m &4$\times$600 s   \\
2009 Jun. 09 &Gemini South/T-ReCS &11.7 $\mu$m &4$\times$1577 s   \\
2009 Apr. 18 &Gemini South/T-ReCS &18 $\mu$m   &2$\times$2065 s   \\
\hline
\end{tabular}\label{tab:newobs}
\end{minipage}\end{center}
\end{table}%\end{center}

A different approach to shed light on the pre-SN evolution of SN~1987A
is to study nearby analogs of SN~1987A's progenitor that have not yet
exploded.  As noted above, three possible cousins of SN~1987A's
progenitor are known: Sher 25, HD 168625, and SBW1.  Of these, SBW1
appears the most similar in terms of the nebular structure and the
luminosity of the central star, but it is not as extensively studied
as the other two.  Smith et al.\ (2007) first discovered SBW1 and
performed the initial study of its nebula.  While it is seen projected
in the Carina Nebula star-forming region, its positive radial velocity
suggests that it is actually located at a much larger distance {\it
  behind} the Carina Nebula; a distance of $\sim 7$\,kpc rather than
the well-established 2.3\,kpc distance to Carina (Smith 2006) also
provides a better match between the expected luminosity from the B1.5
Iab spectral type and the observed magnitude and relatively low
reddening.  This distance makes the luminosity of SBW1 comparable to
that of Sk$-$69$\arcdeg$202, suitable for an 18--25\,M$_{\odot}$
progenitor star.  From our more detailed analysis below, we find that
SBW1 is indeed a virtual twin of the progenitor of SN~1987A and its
circumstellar environment.

\begin{figure}\begin{center}
    \includegraphics[width=2.8in]{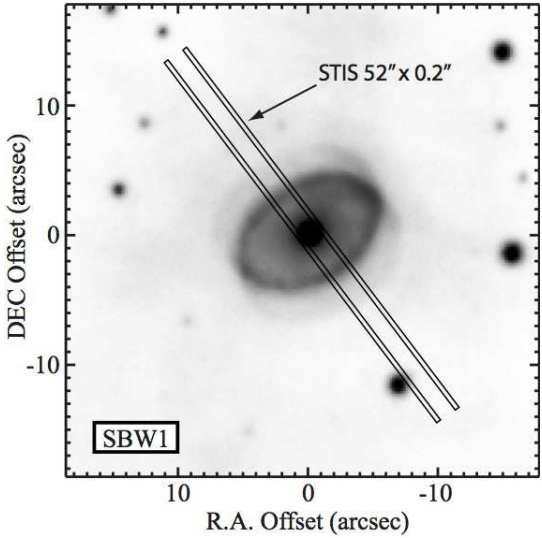}
\end{center}
\caption{Positions of the two STIS long-slit apertures superposed on
  the ground-based H$\alpha$ image of SBW1 from Smith et al.\ (2007).
  The fluxes at these two positions were added to produce the 
  two-dimensional spectrum shown in Figure~\ref{fig:stis}.}
\label{fig:slit}
\end{figure}

\begin{figure}\begin{center}
    \includegraphics[width=2.8in]{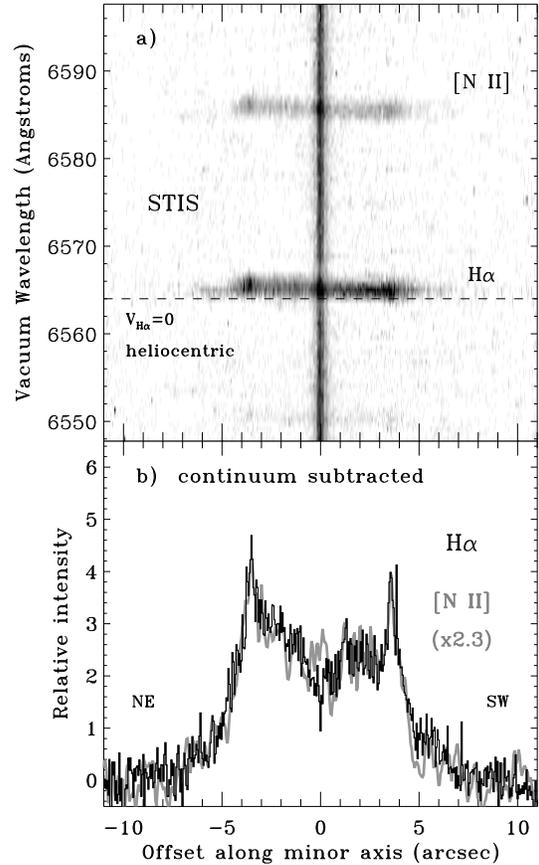}
\end{center}
\caption{Panel (a) shows the two-dimensional STIS spectrum of the SBW1 
  ring centred on H$\alpha$ and [N~{\sc ii}] $\lambda$6584, 
  corresponding to the sum of the fluxes in the two slits shown in
  Figure~\ref{fig:slit}. Panel (b) shows a tracing of the
  continuum-subtracted flux of both H$\alpha$ (black) and [N~{\sc ii}]
  (grey) along the slit.  Positional offset along the slit is shown on
  the horizontal axis in both panels, with northeast (NE) oriented to the
  left.  From the difference in Doppler shift along the slit, it is
  clear that the NE side is the receding side of the equatorial ring.}
\label{fig:stis}
\end{figure}

\section{OBSERVATIONS}

\subsection{HST/WFC3 Images}

Following discovery in our ground-based observations (Smith et al.\
2007), we imaged SBW1 using the newly installed WFC3 camera onboard
{\it HST} on 2009 Dec.\ 8. (UT dates are used throughout this paper; 
see Table~\ref{tab:newobs}.)  We used three
filters: F502N sampling [O~{\sc iii}] $\lambda$5007, F656N sampling
H$\alpha$, and F658N sampling [N~{\sc ii}] $\lambda$6584.  We employed
standard image-reduction techniques, and the resulting monochromatic
F502N, F656N, and F658N images are shown in Figures~\ref{fig:hst}a,
\ref{fig:hst}b, and \ref{fig:hst}c, respectively, whereas a colour
composite of the three is shown in Figure~\ref{fig:hst}d.

The {\it HST}/WFC3 images detect no [O~{\sc iii}] emission from the
ring nebula itself.  This suggests that there is no nearby source of
hard-ultraviolet (UV) photons ($> 35$\,eV) that can ionise 
O$^+$ to O$^{++}$ in the
circumstellar gas.  There are apparently plenty of hard ionising
photons from the much more massive early O-type stars in proximate
regions of the Carina Nebula (also evidenced by the bright diffuse
[O~{\sc iii}] emission seen around SBW1), so this provides yet another
argument that SBW1 is not actually located within the Carina Nebula,
but is instead far {\it behind} it and seen in projection (Smith et
al.\ 2007).  Similarly, we detect no features in absorption associated
with the circumstellar dust in SBW1, contrary to expectations for dust 
features in silhouette against a background screen of an H~{\sc ii} region.

The H$\alpha$ and [N~{\sc ii}] images reveal emission structures that
are essentially identical.  In a F656N $-$ F658N difference image (not
shown), the nebula vanishes almost completely, except for a small
wisp of emission in the diffuse inner part of the ring at the 1\%
level, which could easily be attributable to some foreground
absorption or emission in the Carina Nebula.  As we will see below,
our analysis of the continuum-subtracted emission in {\it HST} Space 
Telescope Imaging Spectrograph (STIS) spectra
shows no variation in the H$\alpha$/[N~{\sc ii}] flux ratio across the
ring, consistent with flux ratios measured in images (STIS spectra are
preferred for this comparison, since scattered starlight can be
subtracted).  Thus, the gas in the dense and thin equatorial ring, the
gas in the outer bipolar regions, and the more diffuse gas filling the
interior of the ring are likely to all have the same relative N/H
abundance.  This is important for some models of the formation of the
nebula (see below).

The most interesting results of the {\it HST} imaging are (1) the
detailed structure in the equatorial ring itself, seen at roughly 7
times the effective {\it spatial} resolution (not angular resolution)
provided by {\it HST} images of the more distant SN~1987A ring, and
(2) diffuse emission structures inside the ring, which are associated
with hot dust features (see below).  Both of these provide important
clues about the structure and origin of the nebula into which the
SN~1987A blast wave has been expanding.  The specific structures and
implications are discussed in \S 3 and \S 4.

\subsection{HST/STIS Spectra}

The SBW1 ring was also observed with {\it HST}/STIS on 2010 May 18
(UT, see Table 1).  The 52\arcsec $\times$ 0$\farcs$2 slit aperture
was oriented along position angle $+$37\arcdeg, at two offset
positions on either side of the central star as shown in
Figure~\ref{fig:slit}.

Figure~\ref{fig:stis}a indicates that there is a slight velocity
gradient across the ring, such that the NE side of the equatorial ring is
redshifted by 10--20\,km\,s$^{-1}$, as was seen in the higher-resolution
echelle spectra of Smith et al.\ (2007).  Also, the entire ring
centroid is redshifted by $\sim 20$\,km\,s$^{-1}$ (the dashed line shows
$v=0$ for H$\alpha$), confirming the systemic-velocity measurement by
Smith et al.\ (2007).  As noted by those authors, this favours a large
Galactic distance on the far side of the Carina spiral arm at 6--7\,kpc 
(similar to well-known supergiants like AG~Car and HR~Car, as well
as the cluster Westerlund 2, etc.), rather than a closer distance that
would put SBW1 inside the Carina Nebula at 2.3\,kpc.

Figure~\ref{fig:stis}a reveals no continuum emission in the interior
parts of the ring.  This means that even though there is dust located
there (as indicated by mid-IR thermal emission; see below), the dust
does not contribute enough scattered light to affect the H$\alpha$
{\it HST}/WFC3 image.  We can therefore assume that gas in the ring's
interior is ionised, and that the corresponding H$\alpha$ emission
measure provides information about the electron density in that region
(the [S~{\sc ii}] lines are detected at too low signal-to-noise ratio in 
our STIS spectra to use their flux ratio as an electron-density diagnostic).

\begin{figure*}\begin{center}
    \includegraphics[width=6.7in]{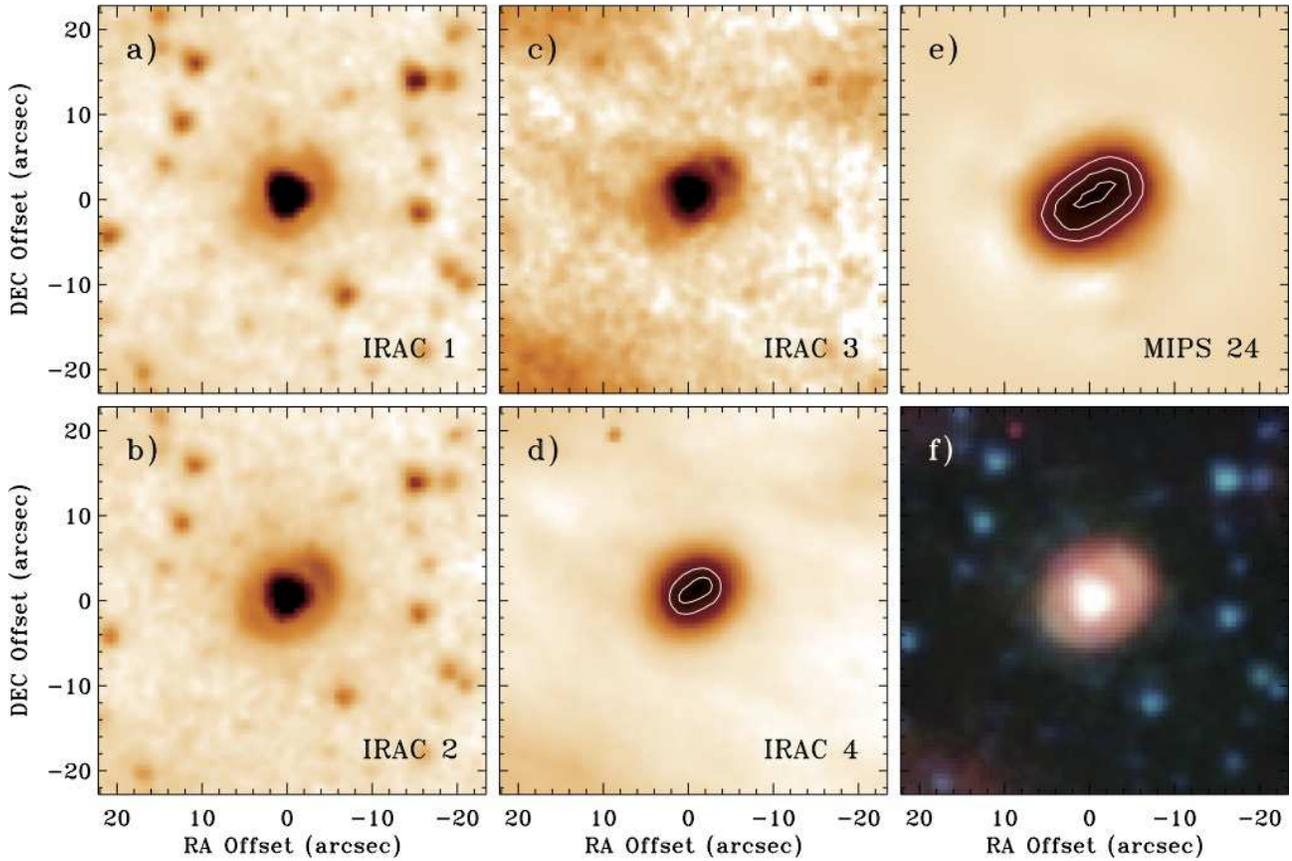}
\end{center}
\caption{IR images of SBW1 taken with {\it Spitzer} in (a) IRAC Band
  1, (b) IRAC Band 2, (c) IRAC Band 3, (d) IRAC Band 4, and (e) MIPS
  24\,$\mu$m.  Panel (f) is a colour composite of IRAC images with
  Band 1 in blue, Band 2 in green, and Band 4 in red.  Contours are
  drawn to show structure near the emission peak in Panels (d) and
  (e).  The 24\,$\mu$m image in Panel (e) is displayed to enhance
  contrast, after having a smoothed version of the image subtracted (a
  partial unsharp mask).  Elongation is apparent in the raw 24\,$\mu$m
  image, however.}
\label{fig:irac}
\end{figure*}

\begin{figure*}\begin{center}
    \includegraphics[width=4.3in]{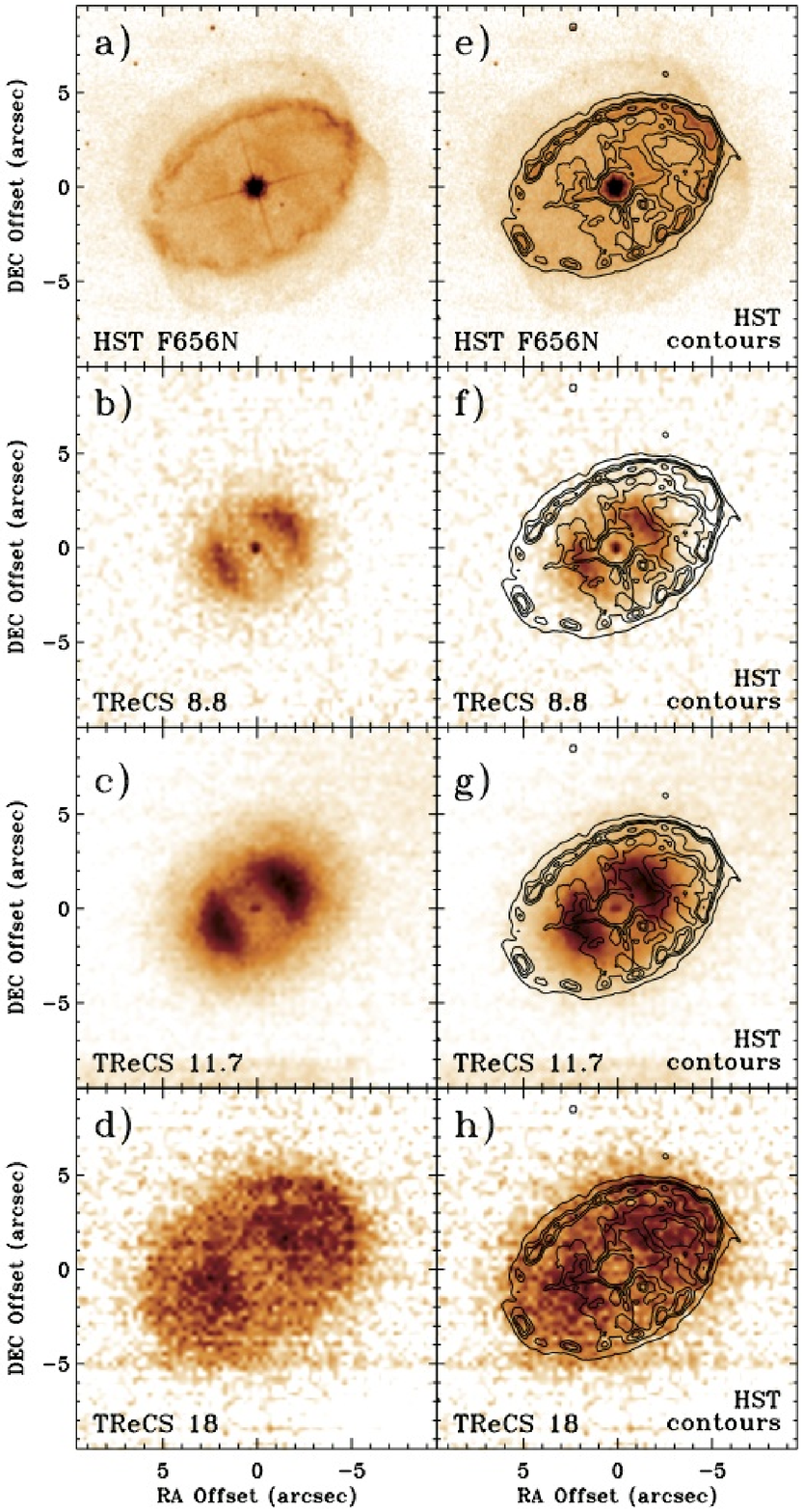}
\end{center}
\caption{Comparison of thermal-IR images taken with Gemini/T-ReCS and
  the {\it HST} image.  The left column [panels (a), (b), (c), and (d)]
  displays the {\it HST}/WFC3 H$\alpha$ image, and the Gemini/T-ReCS images
  at 8.8, 11.7, and 18\,$\mu$m, respectively.  The right column [panels
  (e), (f), (g), and (h)] shows the same images with the contours of
  the {\it HST} image superposed.}
\label{fig:trecs}
\end{figure*}

The continuum-subtracted intensity tracings in Figure~\ref{fig:stis}b
reveal no significant difference in the H$\alpha$/[N~{\sc ii}] flux
ratio across the ring, consistent with WFC3 imaging as noted above.
This suggests that nitrogen abundances and ionisation/excitation
conditions are the same in the dense equatorial ring and in the more
diffuse gas that fills its interior.  It is therefore likely that the
roughly solar values for the nebular N abundance derived by Smith et
al.\ (2007) from ground-based spectra with lower spatial resolution
apply across the entire nebula.

\subsection{Spitzer IRAC and MIPS Images}

The position of SBW1 was observed as part of the {\it Spitzer Space
  Telescope} ({\it Spitzer}) survey of star formation in the Carina
Nebula (P.I.: Smith; see Smith et al.\ 2010b) using both the Infrared
Array Camera (IRAC; Fazio et al.\ 2004) and the Mid-Infrared
Photometer for Spitzer (MIPS; Rieke et al.\ 2004).
Figure~\ref{fig:irac} shows individual images in the four IRAC bands,
the 24\,$\mu$m MIPS image, and a colour composite of the IRAC images
(note that the longer-wavelength 70 and 160\,$\mu$m MIPS images were
saturated). More details about the observations and data reduction are
provided by Smith et al.\ (2010b) and Povich et al.\ (2011).

SBW1 is clearly detected in all five filters observed by {\it
  Spitzer}, and photometry is listed in Table~\ref{tab:sed}.  In IRAC
bands 1--3, the point-like central star dominated the total flux from
SBW1, although low-level extended emission is seen in all three
filters consistent with a few percent of the total flux, and the
relative contribution of the extended emission increases toward longer
wavelengths.  The spatial extent of this extended emission in IRAC
Bands 1--3 is consistent with emission from the dense equatorial ring
seen in {\it HST} images, and the emission mechanism may be a
combination of polycyclic aromatic hydrocarbon (PAH) emission,
free-free, and scattered starlight.  No spectrum of the 3--6\,$\mu$m
IR emission is available, however.

Extending to longer wavelengths in IRAC Band 4 at $\sim$8\,$\mu$m, the
emission is qualitatively different.  The emission is no longer
dominated by an unresolved point source, but instead appears elongated
by a few arseconds in the SE/NW direction, consistent with the major
axis of the nebula.  While clearly not a point source, the Band-4
emission is less spatially extended than the ring emission seen in
Bands 1--3, although the poorer angular resolution of {\it Spitzer} at
$\sim 8$\,$\mu$m does not clearly resolve the structure of the
emitting region.  This $\sim 8$\,$\mu$m emission likely arises from
thermal emission from hot dust that fills the interior regions of the
ring.  This emission is seen more clearly in the Gemini/T-ReCS images
presented below.  It is also evident from an analysis of the spectral
energy distribution (SED) in \S 2.5 that the $\sim 8$\,$\mu$m flux is
dominated by the hot-dust component and not any stellar photospheric
emission, consistent with this interpretation.

The MIPS 24\,$\mu$m image of SBW1 is shown in Figure~\ref{fig:irac}e.
Examining the raw image, it is evident that the source is slightly
elongated along the major axis of the nebula.  We have enhanced the
contrast of axisymmetric structure in this image by subtracting a
smoothed version of the image from the original, and the elongated
nature of the source is clear from the contours drawn in
Figure~\ref{fig:irac}e.  The asymmetry is present on angular scales
larger than the diffraction limit of $\sim 7\arcsec$.  This suggests
that much of the 24\,$\mu$m flux arises from a source larger than the
two peaks of hot-dust emission that dominate the $\sim 8$\,$\mu$m flux
(as seen in IRAC Band 4), since those features are separated by only
$\sim$4{\arcsec}.  Thus, the 24\,$\mu$m flux arises largely from
cooler dust in the outer equatorial ring, and not from the diffuse
emission interior to the ring.  This conclusion is supported by our
analysis of higher-resolution ground-based mid-IR images in the next
section.

\subsection{Gemini South/T-ReCS Images}

We obtained images of SBW1 at 8.8, 11.7, and 18.0\,$\mu$m on 2008
March 30, 2009 April 18 and June 9, and 2010 March 29 using T-ReCS
mounted on the 8\,m Gemini South telescope (see Table 1).  T-ReCS was
the facility mid-IR imager and spectrograph at Gemini South, with a
$320 \times 240$ pixel Si:As IBC array, a pixel scale of 0$\farcs$089,
and a resulting field of view of 28$\farcs$5 $\times$ 21$\farcs$4. The
observations were taken with a 15$\farcs$0 east-west chop throw.
Individual sky-subtracted frames were then combined to make a coadded
image in each filter.  Figure~\ref{fig:trecs} shows the resulting
coadded T-ReCS images at 8.8, 11.7, and 18.0\,$\mu$m, compared to the
{\it HST}/WFC3 H$\alpha$ image on the same scale
(Figure~\ref{fig:trecs}a).  The left column displays the images in
false colour, whereas the right column gives the same images with
contours of the H$\alpha$ emission superposed in order to show the
relative positions of the ionised equatorial ring and the hot inner
dust traced by mid-IR emission.

\begin{figure}\begin{center}
    \includegraphics[width=2.9in]{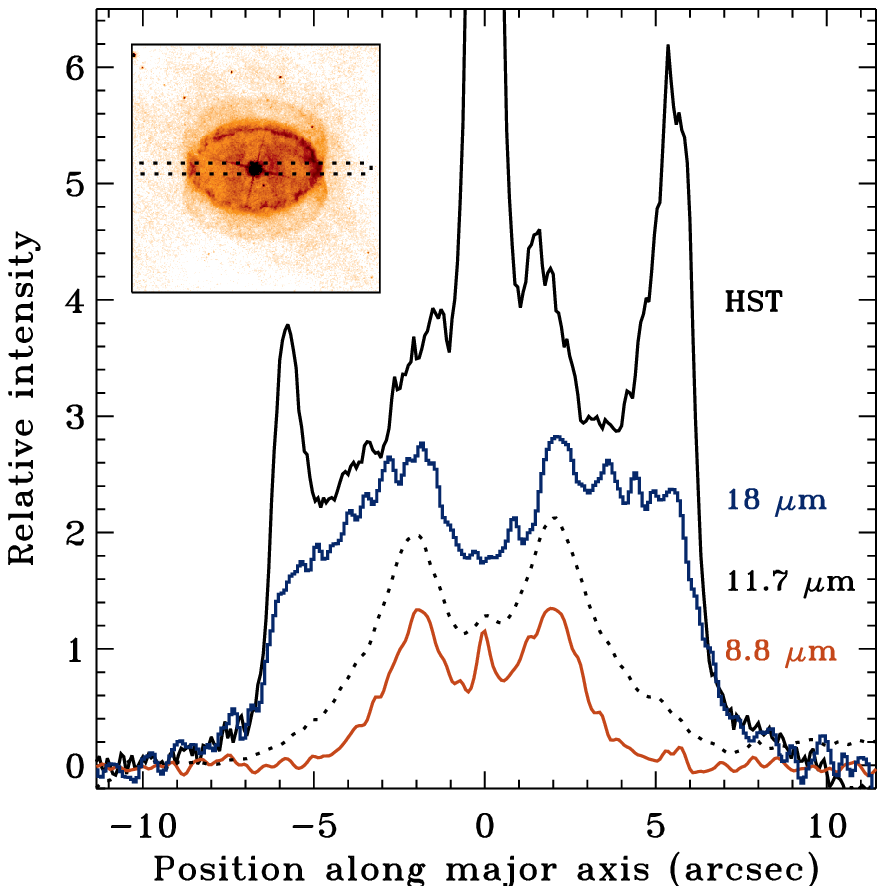}
\end{center}
\caption{Tracings of the relative intensity across the major axis of
  the SBW1 ring in the {\it HST}/WFC3 H$\alpha$ image, and the
  Gemini/T-ReCS images at 8.8 (red), 11.7 (dotted), and 18\,$\mu$m
  (blue).  A dashed box in the inset image indicates the location and
  width of the 1$\farcs$5-wide scan.}
\label{fig:trace}
\end{figure}

A point source at the location of the central star is clearly detected
at 8.8 and 11.7\,$\mu$m.  Both the 8.8 and 11.7\,$\mu$m images were
taken under nonphotometric weather conditions, and the 11.7\,$\mu$m
images were taken over two separate epochs, so we do not use them to
derive {\it absolute} photometry.  However, the resulting images can
be used to provide a precise measurement of the {\it relative}
contribution of the central point source to the total flux in each
filter.  We find that the central star contributes 2.5\% $\pm 0.2$\%
of the total 8.8\,$\mu$m flux measured in a 5{\arcsec}-radius circular
aperture, and similarly, the central star contributes 0.6\% $\pm
0.08$\% of the total 11.7\,$\mu$m flux measured in a 7{\arcsec}-radius 
circular aperture (a larger aperture was used to measure the
11.7\,$\mu$m total flux because the 11.7\,$\mu$m emission is more
extended, with some contribution from the main equatorial ring).  The
central star is not detected in the 18\,$\mu$m filter.  The fractions
of the total flux contributed by the central point source are useful
for our analysis of the SED discussed below (\S 2.5).

The T-ReCS images also provide critical information about the spatial
distribution of warm dust grains in the SBW1 nebula, which is unclear
from the lower-resolution imaging with {\it Spitzer}.  After
separating out the central star, the spatially resolved extended
structure seen in the mid-IR can be understood as two spatial
components whose relative contribution to the total flux changes with
wavelength: (1) emission from the thin equatorial ring with a
semimajor axis of $\sim 6${\arcsec}, whose emission becomes relatively
stronger with increasing wavelength, and (2) diffuse emission arising
from dust distributed throughout the interior of the ring, but
concentrated mainly in two arcs of emission located
2{\arcsec}--3{\arcsec} to the SE and to the NW of the central star,
whose contribution to the total flux decreases with increasing
wavelength.  Both the thin equatorial ring and the inner diffuse
emission can be seen in the H$\alpha$ image taken with {\it
  HST}. Comparing this H$\alpha$ emission to the T-ReCS images (right
column of Figure~\ref{fig:trecs}), it is evident that the equatorial
ring contributes no detectable emission to the 8.8\,$\mu$m image.  The
outer equatorial ring emission can be seen as a faint halo in the 
11.7\,$\mu$m image, as well as the outer boundary of the (noisy) 18\,$\mu$m
image.  This suggests that the equatorial ring contains relatively cool
dust.  The inner double-peaked structure associated with the more
diffuse emission inside the ring dominates the 8.8 and 11.7\,$\mu$m
images, but is less prominent at 18\,$\mu$m.  This suggests that the
inner dust is relatively hot.

Tracings of the intensity along the major axis of SBW1 are shown in
Figure~\ref{fig:trace}, comparing the {\it HST} H$\alpha$ emission to
the mid-IR emission in the three filters observed with T-ReCS at
Gemini South (the intensities are scaled arbitrarily for display).
These tracings confirm the impression from images that the the 8.8 and
11.7\,$\mu$m flux is dominated by two inner peaks of emission located
within $\pm 3${\arcsec} from the star.  The location of the peak of
this emission and the profile shape in tracings at 8.8 and 11.7\,$\mu$m
match, indicating that there is no strong radial temperature gradient
in the dust.  It is also clear from the tracings in
Figure~\ref{fig:trace} that these two peaks of hot-dust emission
coincide with a subtle enhancement of H$\alpha$ emission.  The 18\,$\mu$m 
emission has a flatter distribution along the major axis of the
nebula, with the flux decreasing only slightly from
2{\arcsec}--6{\arcsec} away from the star.  Like the equatorial ring
seen in H$\alpha$ by {\it HST}, the 18\,$\mu$m emission drops off
abruptly at $\sim$ 6$\farcs$5 from the star (Figure~\ref{fig:trace}),
providing strong evidence that cooler dust in the ring emits a
considerable fraction of the $\sim 20$\,$\mu$m emission.  Tracings at
all three mid-IR wavelengths show a pronounced deficit of emission
within 2{\arcsec} of the central star, revealing an inner region
devoid of dust grains.

\begin{table}\begin{center}\begin{minipage}{3.3in}
      \caption{Adopted IR flux densities of SBW1}
\scriptsize
\begin{tabular}{@{}lccc}\hline\hline
Tel./Instr.   &Filter/$\lambda$  &$F_{\nu}$ &$\pm$ \\
...           &(name or $\mu$m)  &(Jy)     &(Jy) \\ \hline
2MASS         & $J$ / 1.235 $\mu$m  &0.0959   &0.0020  \\
2MASS         & $H$ / 1.662 $\mu$m  &0.0866   &0.0019  \\
2MASS         & $K$ / 2.159 $\mu$m  &0.0697   &0.0014  \\
WISE          &3.4 $\mu$m        &0.0369   &0.0014  \\
WISE          &4.6 $\mu$m        &0.0256   &0.0012  \\
WISE          &12 $\mu$m         &0.815    &0.038   \\
WISE          &22 $\mu$m         &6.89     &0.32    \\
MSX           &A / 8.28 $\mu$m   &0.3034   &0.0149  \\
MSX           &C / 12.13 $\mu$m  &1.328    &0.0969  \\
MSX           &D / 14.65 $\mu$m  &1.710    &0.115   \\
MSX           &E / 21.34 $\mu$m  &6.415    &0.398   \\
AKARI         &9 $\mu$m          &0.3102   &0.0086  \\
AKARI         &18 $\mu$m         &7.849    &0.484   \\
Spitzer/IRAC  &3.6 $\mu$m        &0.039    &0.0015  \\
Spitzer/IRAC  &4.5 $\mu$m        &0.032    &0.0016  \\
Spitzer/IRAC  &5.8 $\mu$m        &0.027    &0.0032  \\
Spitzer/IRAC  &8.0 $\mu$m        &0.209    &0.0229  \\
Spitzer/MIPS  &24 $\mu$m         &7.688    &1.23    \\
\hline
\end{tabular}\label{tab:sed}
\end{minipage}\end{center}
\end{table}%\end{center}

\begin{figure}\begin{center}
 \includegraphics[width=3.3in]{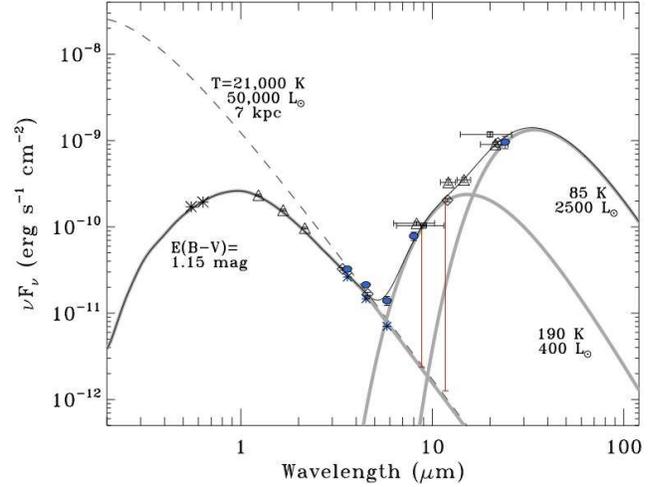}
\end{center}
\caption{The optical/IR SED of SBW1.
  The optical $V$ and $R$ magnitudes (asterisks) are from Table 1 of
  Smith et al.\ (2007).  $JHK$ magnitudes from 2MASS are shown with
  unfilled triangles. Unfilled diamonds, triangles, and squares show
  catalog photometry from WISE (3.4, 4.6, 12, and 22\,$\mu$m), MSX
  (8.3, 12.1, 14.7, and 21.3\,$\mu$m), and AKARI (9 and 18\,$\mu$m),
  respectively.  These all represent the spatially unresolved total
  flux of the star and nebula.  The filled blue dots represent
  photometry of the entire object from our {\it Spitzer} IRAC and MIPS
  data, whereas the blue asterisks represent the measured flux of the
  central star in the IRAC images in Bands 1--3.  Finally, the orange
  vertical bars represent the flux of the central star in
  our 8.8 and 11.7\,$\mu$m T-ReCS images, where we measured the
  fraction of the total flux contributed by the central point source
  and then scaled the total flux to match the SED (the images were
  obtained in some nonphotometric conditions). The dashed curve shows a
  $T =$ 21,000\,K blackbody, representative of the unreddened B1.5~Iab 
  supergiant.  The three thick grey curves are this stellar
  blackbody reddened by $E(B-V) = 1.15$\,mag, plus two grey bodies
  (emissivity $\propto \lambda^{-1}$) representing thermal emission
  from dust at approximately 190\,K and 85\,K.  The solid black curve is
  the total flux contributed by the sum of these three components.}
\label{fig:sed}
\end{figure}

\subsection{The SED}

Figure~\ref{fig:sed} shows the SED of SBW1 from optical through IR
wavelengths using data from several sources.  We obtained
publically available IR photometry of SBW1 from the Infrared Science
Archive (IRSA\footnote{\tt
  http://irsa.ipac.caltech.edu/Missions/missions.html}), including
measurements from 2MASS (Skrutskie et al.\ 2006), and from 
point-source catalogs of the Midcourse Space Experiment (MSX; Price 1995),
Wide-field Infrared Survey Explorer (WISE; Wright et al.\ 2010), and
AKARI satellites.  Flux densities (in Jy) from these sources are
listed in Table~\ref{tab:sed}, and then converted to $\nu F_{\nu}$ in
Figure~\ref{fig:sed}.  We also show $V$- and $R$-band photometry from
Smith et al.\ (2007) in Figure~\ref{fig:sed}.

Figure~\ref{fig:sed} and Table~\ref{tab:sed} include photometry
measured from our {\it Spitzer} IRAC and MIPS surveys of the Carina
Nebula (Smith et al.\ 2010b).  Filled blue circles in
Figure~\ref{fig:sed} correspond to total fluxes of SBW1 measured in a
12{\arcsec}-radius circular aperture in all 5 bands, while blue
asterisks are the stellar flux in Bands 1, 2, and 3 measured in a
3{\arcsec}-radius aperture with corrections for point-spread function
(PSF) flux outside this radius.  

Lastly, the long vertical red/orange bars in Figure~\ref{fig:sed} are
derived from the fraction of the total flux contributed by the
central star in our 8.8 and 11.7\,$\mu$m T-ReCS images of SBW1 from
Gemini South.  The upper end of these bars is tied to the solid-curve
fit to the SED (see below), and the bottom of each then represents the
expected stellar flux at each wavelength.

As shown in Figure~\ref{fig:sed}, the observed optical and IR
photometry of SBW1 can be approximated reasonably well with three
simple emission components: (1) a 21,000\,K blackbody corresponding to
the B1.5~Iab central star (dashed curve), reddened by foreground
extinction of $E(B-V) = 1.15$\,mag, (2) a greybody (with an emissivity
proportional to $\lambda^{-1}$ at long wavelengths) representing warm
dust at 190\,K, and (3) another greybody representing cooler dust at 85\,K.  
At a distance of 7\,kpc, these stellar-photosphere, hot-dust, and
cool-dust components have luminosities of 50,000\,L$_{\odot}$,
400\,L$_{\odot}$, and 2500\,L$_{\odot}$, respectively.

Photometric information for the unresolved central star is consistent
with reddened photospheric emission at all wavelengths up to $\sim 
12$\,$\mu$m.  There is no evidence for thermal free-free emission from an
extended wind photosphere (e.g., Wright \& Barlow 1979) at these
wavelengths, implying a rather weak stellar wind from this B~1.5~Iab
supergiant.  The uncertainty suggests that the free-free stellar wind
emission at $\sim 10$\,$\mu$m is less than about 10\% of the stellar
flux, which in turn is 0.5--1\% of the total observed flux from
SBW1.  This would correspond to a flux density of roughly 0.5--1\,mJy
or less at 10\,$\mu$m for any stellar-wind emission.  Using this IR flux
and a distance of 7\,kpc in Equation (45) of Puls et al.\ (2008), we
find a likely upper limit to the central star's mass-loss rate of

\begin{equation}
  \dot{M} < 4 \times 10^{-7} \big{(}v_{300}\big{)} \big{(}F_{10}D_7\big{)}^{3/4} \ {\rm M}_{\odot}\,{\rm yr}^{-1},
\end{equation}

\noindent 
where $v_{300}$ is the terminal speed of the BSG wind in units of
300\,km\,s$^{-1}$, $F_{10}$ is the 10\,$\mu$m flux density in mJy (the
relevant flux is 0.5--1 mJy), and $D_7$ is the distance relative to
7\,kpc.  This upper limit of $\dot{M} <$ (2--4) $\times
10^{-7}$\,M$_{\odot}$\,yr$^{-1}$ for the mass-loss rate of the central
blue supergiant star of SBW1 could be a factor of a few lower if the
likely effects of clumping are included.  In any case, this upper
limit is in good agreement with upper limits to the mass-loss rate of
SN~1987A's progenitor of (1.5--3) $\times
10^{-7}$\,M$_{\odot}$\,yr$^{-1}$ inferred from hydrodynamic
simulations of the nebula (Blondin \& Lundqvist 1993; Martin \& Arnett
1995) and an estimated mass-loss rate of $7.5 \times 
10^{-8}$\,M$_{\odot}$\,yr$^{-1}$ inferred from the observed 
expansion rate of the radio photosphere (Chevalier \& Dwarkadas 1995).

The sharp increase in total flux from 6 to 8\,$\mu$m, as the SED
transitions from a photosphere to warm-dust emission, is indicative of
a cavity with little or no dust close to the star.  The 190\,K ``hot''-dust 
component dominates the total flux from about 6 to 15\,$\mu$m, and
arises from the diffuse structures filling the interior of the ring,
as seen in the double-peaked feature in the 8.8 and 11.7\,$\mu$m images
from Gemini South (Figure~\ref{fig:trecs}).  The cooler 85\,K dust
component dominates the total flux longward of about 18\,$\mu$m, and
this cooler dust appears to reside primarily in the equatorial ring.
We cannot rule out the presence of some cooler dust that may produce
excess far-IR luminosity (and much higher mass) from presently
available data.

One can calculate the approximate mass of dust grains required to emit
each of these two components by making some simplifying
assumptions about the grain-emissivity properties, and taking the
mid-IR emission to be optically thin. Following Smith et al.\ (2003),
the total mass of emitting dust can be expressed as

\begin{equation}
  M_d = \frac{4 D^2 \rho \, (\lambda F_{\lambda})}{3 \, 
(\lambda Q_{\rm e} / a) \, B_{\lambda}(T) },
\end{equation}

\noindent 
where ($\lambda Q_{\rm e} / a$) is a quantity that describes the
grain-emission efficiency $Q_{\rm e}$ (see Draine \& Lee 1984).  We assume a
distance $D=7$\,kpc for SBW1, and astronomical silicate with a typical
grain density $\rho = 3$\,g\,cm$^{-3}$ (at these wavelengths the assumed
grain radius is not critical as long as the typical grains are less
than $a = 5$\,$\mu$m).  With these parameters we derive dust masses of
$M_d(190) = 1.3 \times 10^{-5}$\,M$_{\odot}$ and $M_d(85) =
4.7 \times 10^{-3}$\,M$_{\odot}$ for the $T_d = 190$\,K and 85\,K
components of the SED in Figure~\ref{fig:sed}, respectively.

The typical uncertainty for this type of rough estimate of the dust mass
is $\pm 30$\%, dominated by assumptions about grain properties, as well
as uncertainties in the ranges of temperatures that can fit the SED.
We have assumed astronomical silicate as the nominal dust composition,
which seems the most reasonable given that SBW1 is a massive evolved
star with normal CNO chemical abundances (Smith et al.\ 2007).
Moreover, IR spectra of SN~1987A obtained with {\it Spitzer} revealed
strong silicate emission features (Bouchet et al.\ 2006).
Unfortunately, we do not have direct observational constraints on the
grain composition in SBW1.  Had we instead adopted the assumption of
small ($a = 0.2$\,$\mu$m) graphite grains and calculated the dust mass
following the method described by Smith \& Gehrz (2005), we would have
derived dust masses of $M_d(190) = 1.1 \times 10^{-5}$\,M$_{\odot}$ and 
$M_d(85) = 8.7 \times 10^{-3}$\,M$_{\odot}$ for the
$T_d = 190$\,K and 85\,K components, respectively.  The mass for the
warmer component agrees within the expected uncertainty with that
derived above for silicates.  The cooler component is a factor of
$\sim 2$ larger for carbon grains, but as noted above, carbon grain
composition seems unlikely since SBW1 is not C-enriched.  In any case,
we regard the value of $M_d(85) = 4.7 \times 10^{-3}$\,M$_{\odot}$ as
a lower limit to the possible dust mass, since our observations do not
constrain the SED longward of 24\,$\mu$m, making it possible that a
larger mass of cooler ($T_d < 85$\,K) dust might reside in the
equatorial ring, or just outside it.

To extend this dust mass to a total nebular mass requires an
assumption of the gas-to-dust mass ratio, which is poorly constrained but
generally taken to be about 100:1 in massive-star nebulae (see, e.g.,
Smith et al. 2003, and references therein).  We thus infer a total gas
mass of order 0.5--1\,M$_{\odot}$ for the ring nebula around SBW1,
derived from the mass of the coolest dust we detect.  Although the
gas:dust ratio could be substantially larger for the hotter 190\,K dust
component that is mixed with ionised gas in the interior region of the
ring, this component makes a negligible contribution to the total
nebular mass.  Interestingly, a mass of order 0.5--1\,M$_{\odot}$ for
the SBW1 ring nebula is about equal to the mass of pre-SN ejecta
surrounding SN~1987A.  The directly observed mass of ionised gas in
the SN~1987A ring is at least 0.04\,M$_{\odot}$, although this is a
lower limit to the nebula mass because it is derived from emission
lines and corresponds only to a thin skin at the inward-facing edge of
the ring that was ionised by the SN; observations of light echoes
reveal a much larger mass of material in the ring and outside it of
$\sim 1.7$\,M$_{\odot}$ (Sugerman et al.\ 2005).\footnote{Note that the
  lower dust mass and higher dust temperature derived from recent
  observations of the SN~1987A ring by Bouchet et al.\ (2006)
  correspond to a very different physical regime of dusty gas that is
  heated and partly destroyed by the SN shock, so these do not
  necessarily reflect the properties of dust originally surrounding
  the progenitor.}

The cool-dust component that dominates the $\lambda > 20\,\mu$m
emission has a total luminosity that is $\sim 5$\% of the original
stellar luminosity.  The vertical thickness of the dust in the
equatorial ring must therefore be $\sim 10$\% of its radius, or
$6 \times 10^{16}$\,cm, based on the fraction of the total luminosity
that it intercepts.  The much lower luminosity of the hot-dust
component, despite its closer separation from the central star, means
that it is very optically thin, and so its relative luminosity cannot
provide a meaningful constraint on the vertical thickness of this 
hot-dust component.

\subsection{X-ray Emission}

The Carina Nebula was observed as part of a large program with the
{\it Chandra X-ray Observatory}, called the {\it Chandra} Carina
Complex Project (Townsley et al.\ 2011).  There is no X-ray source
listed in the resulting catalog of X-ray sources in this survey (Broos
et al.\ 2011) within a radius of 15{\arcsec} from SBW1's position.
The quoted completeness limit of this survey would suggest an
absorption-corrected upper limit to the X-ray luminosity of
$10^{30.7}$\,erg\,s$^{-2}$ for sources within the Carina nebula
region. It is, however, difficult to place a meaningful upper limit on
the intrinsic soft X-ray luminosity of SBW1, since it is located at a
distance of 7\,kpc, far behind the Carina nebula ($D=2.3$\,kpc; Smith
2006).  Thus, there could be a large absorption column, and its
intrinsic soft X-ray luminosity could be substantially higher than
this nominal upper limit.

\begin{figure}\begin{center}
    \includegraphics[width=3.2in]{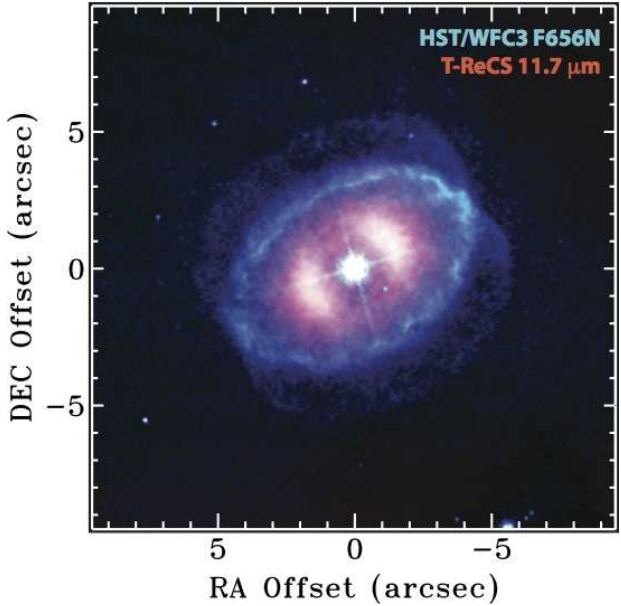}
\end{center}
\caption{A colour composite of the {\it HST}/WFC3 F656N image from
  Figure~\ref{fig:trecs}a displayed in blue/green, and the T-ReCS 
  11.7\,$\mu$m image from Figure~\ref{fig:trecs}c displayed in 
  red/orange.}
\label{fig:color}
\end{figure}

\begin{figure*}\begin{center}
    \includegraphics[width=6.9in]{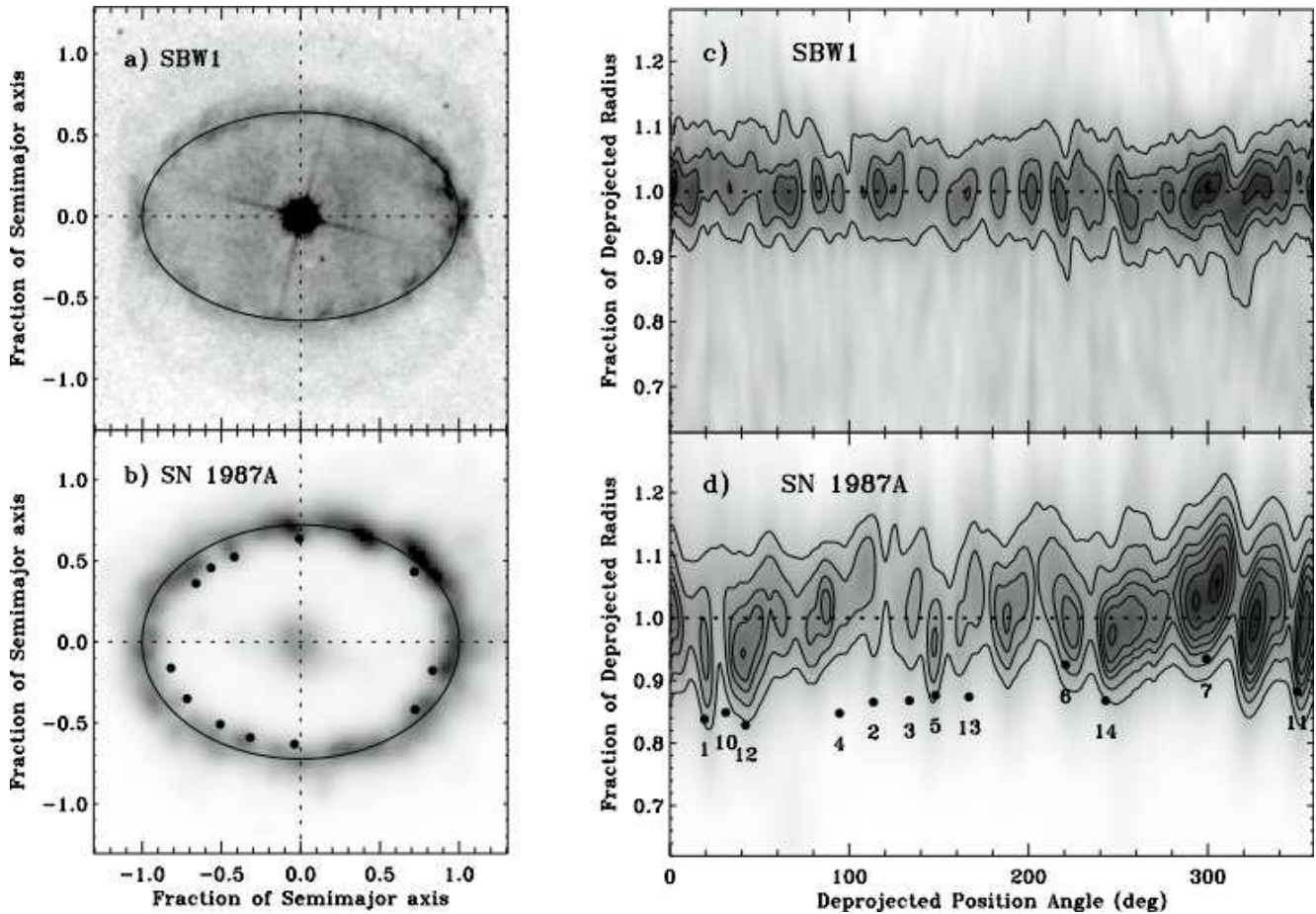}
\end{center}
\caption{Panels (a) and (b) show H$\alpha$ images of the equatorial
  rings surrounding SBW1 and SN~1987A, with best-fit ellipses drawn
  over the images, and rotated so that the major axes are horizontal.
  The {\it HST} image of SN~1987A is from Sugerman et al.\ (2002), and
  the black dots in panel (b) mark the locations of hotspots from
  Table 2 of Sugerman et al.\ (2002).  Panels (c) and (d) show
  the observed H$\alpha$ intensity in terms of deprojected radius from
  the central object, plotted against the deprojected position angle.
  Panel (d) is essentially the same as Figure~7c in Sugerman et al.\
  (2002), and panel (c) is plotted the same way for comparison.}
\label{fig:deproject}
\end{figure*}

\section{OBSERVED MULTI-WAVELENGTH STRUCTURE}

Figure~\ref{fig:color} shows the relative spatial distribution of
ionised gas and warm dust in colour, with the H$\alpha$ image tinted as
blue/green and the mid-IR emission tinted as red/orange. This image
encapsulates the basic multiwavelength structure of SBW1, and is
useful in the discussion below.  One of the curious things about
Figure~\ref{fig:color} is that it runs counter to normal expectations
that dusty regions will be farther from the source of radiation than
the ionised gas.  This and other issues are clarified below.

\subsection{Detailed Structure of the Dense Equatorial Ring}

The new {\it HST}/WFC3 images in H$\alpha$ and [N~{\sc ii}] provide a
detailed view of the morphology in the ring nebula around SBW1. To the
extent that SBW1 is a suitable analog, these images also provide our
best view of the circumstellar environment around an object like
SN~1987A.  The most prominent feature in the images is a thin ring of
emission, presumably in the equatorial plane.  The new {\it HST}
images also provide a better view of the outer rings or hourglass
structure in SBW1, and they clarify the nature of the diffuse emission
structures in the ring's interior, closer to the central star.  Below
we examine the structure of the ring in detail, and then conduct a
comparison with the equatorial ring of SN~1987A.  Structures inside
the ring are discussed in the following section.

In general, the equatorial ring around SBW1 appears as a fragmented
chain of clumps or filaments with a thin radial extent.  Filaments in
the ring are marginally resolved with a thickness of $\sim$ 0$\farcs$1,
or 0.0034\,pc at a distance of 7\,kpc. By assuming this as a typical
thickness for the ionised emitting layer, we can provide a rough
estimate of the density of ionised gas in the ring from the H$\alpha$
emission measure, $EM \, = \, \int \, n_e^2 \, dl$.  This can be
conveniently expressed (see Smith et al.\ 2010a) as

\begin{equation}
n_e \, = \, 15.0 \, \sqrt{\frac{I_{{\rm H}\alpha}}{L_{\rm pc}f}} \ \ {\rm cm}^{-3},
\end{equation}

\noindent 
where $I_{{\rm H}\alpha}$ is the H$\alpha$ line intensity measured in
our narrow-band F656N WFC3 image in units of
$10^{-15}$\,erg\,s$^{-1}$\,cm$^{-2}$\,arcsec$^{-2}$, $L_{\rm pc}$ is
the emitting path-length through the filament in pc, and $f$ is a
geometric filling factor.  Based on the clumpy structure in images, we
adopt $f = 0.5$.  Although this assumption dominates the uncertainty,
it is difficult to quantify.  We adopt a $\pm 25$\% uncertainty in the
value of $f$, which translates to a $\pm 11$\% uncertainty in the
resulting value of $n_e$.  In our F656N image (note that the F656N
filter on WFC3 includes only H$\alpha$, and not [N~{\sc ii}]
$\lambda$6583), we measure an intensity of $6.70 \times
10^{-17}$\,erg\,s$^{-1}$\,cm$^{-2}$\,arcsec$^{-2}$, which translates
to $9.7 \times 10^{-16}$\,erg\,s$^{-1}$\,cm$^{-2}$\,arcsec$^{-2}$
after correcting for $E(B-V) = 1.15$\,mag (see Figure~\ref{fig:sed}).
With these values, we find $n_e = 370 \pm 40$\,cm$^{-3}$.  As an
independent estimate, from the [S~{\sc ii}] $\lambda\lambda$6717, 6731
flux ratio measured in ground-based spectra we found $n_e \approx
500$\,cm$^{-3}$ (Smith et al.\ 2007), with a likely uncertainty of
$\pm 20$\%.  These two estimates are not too discrepant, given the
uncertainties.  By assuming that the toroidal geometry of the ring is
filled with the average of these two density estimates, we would
derive a likely mass of emitting ionised gas of 0.012\,M$_{\odot}$
(see Smith et al.\ 2007).

This estimate of the mass of ionised gas is much less than the
expected total mass of H gas.  The ring's dust mass of
$5 \times 10^{-3}$\,M$_{\odot}$ measured from the luminosity of
$\sim 85$\,K dust in the SED (see \S 2.5; Figure~\ref{fig:sed}) would
imply a total gas mass of roughly 0.5\,M$_{\odot}$ if the equatorial
ring has a normal gas:dust mass ratio of 100:1.  This implies that the
ionisation fraction of the ring is only a few per cent.  The
equatorial ring is therefore ionisation bounded, with the location of
the ionisation front determined by the flux of ionising photons from
the central star as well as the initial density structure of the
ring.  The incomplete ionisation of the ring has a strong impact on
the shape of the rest of the nebula around SBW1, and we return to the
issue in \S 4.1 and \S 4.2.

A strong increase in the ionising flux would be expected to ionise a
larger fraction of the total mass in the ring, and would affect the
apparent-brightness structure of the ring.  This may account for some
of the observed morphological differences between SBW1 and the ring of
SN~1987A, because the ring around SN~1987A is viewed after a huge
burst of ionising radiation from the UV flash of the SN.  For example,
currently the diffuse gas that fills the interior of the SBW1 ring is
bright compared to the thin ring itself, making SBW1 appear more
filled-in than SN~1987A (see Figure~\ref{fig:deproject}).  If the
ionising flux of SBW1 were to increase suddenly, we would expect the
thin equatorial ring to get much brighter, because more of the
high-density neutral gas in the ring would be ionised and the emission
measure (proportional to $n_e^2$) would rise, whereas the interior of
the ring would not brighten because it is already fully ionised.
Given the difference in ionising flux, however, the morphologies of
the two rings are already strikingly similar.

Figure~\ref{fig:deproject} shows H$\alpha$ images of the equatorial
rings around both SBW1 and SN~1987A, as well as plots of the
deprojected intensity around the equatorial ring for each source.
These were produced by first fitting an ellipse to the {\it HST} image
of each ring, and then ``deprojecting'' the appearance of the source
by stretching the image by the appropriate factor along the minor
axis, in order to make that ellipse into a circle.  The deprojected
intensity plots in Figure~\ref{fig:deproject} are then a radial
tracing of the intensity at all position angles around the ring.  For
SN~1987A, this is meant to duplicate the analysis by Sugerman et al.\
(2002), which provides a basis for comparison.  We then performed the
same analysis for SBW1.  For SN~1987A we adopted a ring inclination
of $i = 43\fd$8 (Sugerman et al.\ 2002), and for SBW1 we
measured $i = 50\fd$6 $\pm 0\fd$6, in agreement with an earlier
estimate of $i = 50\fd$2 $\pm 1${\arcdeg} from a ground-based image
(Smith et al.\ 2007).

Figures~\ref{fig:deproject}c and \ref{fig:deproject}d are surprisingly
similar.  If we characterise the ring structure as dominated by a
number of beads along a string, we see that both the wiggles in the
string (${\Delta r}/R$) and rough angular scale (${\Delta l}/R$) of the
beads are similar in the two objects, where $R$ is the average ring
radius (0.21\,pc for SN~1987A and 0.19\,pc for SBW1; see Smith et al.\
2007).  The small variations in $\Delta r$ are typically about $\pm 4$\%
of $R$ for SBW1 and $\pm 6$\% of $R$ for SN~1987A, with a few features
reaching inward to 85\% of $R$ for both objects.  The angular scale of
clumps appears somewhat larger for SN~1987A in these figures, but this
depends on the effective spatial resolution, which is slightly higher for
SBW1 because it is closer (we have smoothed the SBW1 image by about a
factor of 3, but it is a factor of 7 closer).  A typical separation between
major clumps in the ring is 8{\arcdeg} to 15{\arcdeg} for both, with
roughly 22 to 25 major clumps around each ring.  Note that these
figures are intended to exaggerate the differences in structure
between the two rings, with only a portion of $R$ being plotted.

The main interesting result of this analysis is that if we allow for
some small differences in spatial resolution and the precise
arrangement of blobs, the equatorial ring around SBW1 is for all
practical purposes indistinguishable from that around SN~1987A, in
terms of its basic physical attributes.  Yet, SBW1 is a factor of 7 closer
to us, and so in the original {\it HST} image we have 7 times better
effective spatial resolution and can see more details of the structure
than for SN~1987A.  The chief reason this is interesting is that
SBW1 {\it does not exhibit the long R-T fingers} envisioned for
SN~1987A in order to explain the occurrence of hotspots when the tips
of the R-T fingers were hit by the blast wave.  Instead, SBW1 appears
to be a fragmented, clumpy ring with a small radial extent.  Since R-T
fingers are expected to mark the contact discontinuity between the
faster BSG wind and the slower and flattened RSG disk wind, the lack
of such features carries implications for the formation mechanism of
the ring: the ring is probably not formed by a fast BSG wind
sweeping into an extended disk.  Indeed, after analysing the
structures interior to the ring in the next section, we conclude that
the expanding BSG wind is not directly interacting with the dense
equatorial ring, due to the increased pressure resulting from
photoionisation of the ring.  The clumpy dense ring must therefore
have had a different origin, as we discuss later.

Lastly, we note a possibly interesting feature of the ring morphology.
The brightest part of the ring on the north-west edge, at position angle
P.A. $\approx 310${\arcdeg}, coincides with a pronounced kink in the
ring, exhibiting the largest jump in deviations from the average ring
radius (see Figure~\ref{fig:deproject}c).  The significance or cause
of this structure is not obvious. However, a similar kink is seen to
be associated with the brightest portions of the ring around SN~1987A
as well, coincidentally also located at P.A. $\approx 310${\arcdeg}. One
can readily imagine how a single large departure from azimuthal
symmetry could arise in a sudden ejection in a binary system, but this
is harder to explain when the structure of the ring arises from
colliding winds alone (this is discussed in more detail below).

\subsection{Structures Inside the Ring}

The most unexpected result of this study of SBW1 concerns the diffuse
emission from structures in the interior of the equatorial ring.
While ground-based H$\alpha$ images revealed some diffuse emission
apparently filling the ring interior (Smith et al.\ 2007), it was
difficult to reliably separate contributions from the ring and the
stellar PSF.  The new {\it HST}/WFC3 F656N image shows that the
distribution is not uniform across the ring; there are two regions of
enhanced H$\alpha$ emission located about 1$\arcsec$ to 3$\arcsec$ on
either side of the central star, and there is a marked deficit of
H$\alpha$ emission within 1$\arcsec$ to 2$\arcsec$ of the star (see
Figures~\ref{fig:hst} and \ref{fig:trecs}).

More striking, however, are the high-resolution mid-IR images obtained
with T-ReCS on Gemini South.  In the mid-IR, we see two distinct peaks
of warm-dust emission located 1$\arcsec$ to 3$\arcsec$ on either side
of the central star, oriented along the minor axis of the ring (see
Figures~\ref{fig:trecs} and \ref{fig:color}).  These dust-emission
peaks appear to be spatially coincident with the enhancements of
H$\alpha$ mentioned above, but have a much stronger contrast in the
mid-IR.  The two peaks are probably caused by limb brightening of a
toroidal structure with a projected radius of $\sim 2\arcsec$, and if
they are located in the equatorial plane, these inner dust peaks
occur at about 1/3 of the radius of the dense equatorial ring (which
is $\sim 0.2$\,pc; Smith et al.\ 2007).  The IR emission also shows a
clear deficit within a few arcseconds of the star, indicating an inner
region that is relatively devoid of hot dust.

The double-peaked mid-IR emission has the strongest contrast at the
shorter mid-IR wavelengths (8.8 and 11.3\,$\mu$m), whereas the emission
structure appears more uniform at the longest thermal-IR wavelength
(18\,$\mu$m) due to a rising contribution from the outer equatorial
ring.  This suggests that the double-peaked features have the hottest
dust, consistent with their location closer to the star than the other
structures.  Analysis of the optical/IR SED of SBW1
(Figure~\ref{fig:sed}; \S 2.5) suggests the presence of two dust
components emitting at $\sim 190$\,K and 85\,K.  The mid-IR images
therefore indicate that the warmer 190\,K component must be associated
with the inner double-peaked emission, and that the cooler 85\,K
component is associated with cooler dust located in the dense
equatorial ring.

That the double-peaked emission inside the ring is seen in both
H$\alpha$ and in mid-IR continuum emission suggests that the hot dust
is intermixed with ionised gas.  In that case, the heating of the dust
could be collisions with dust, trapped Ly$\alpha$ radiation, or direct
heating by stellar radiation.  The equilibrium grain temperature due
only to stellar radiation is given by

\begin{equation}
  T_g = 28 \Big{[} \frac{Q_{\rm abs}}{Q_{\rm e}} \frac{L}{10^4\,{\rm L}_{\odot}} \big{(}\frac{R}{10^4\,{\rm AU}}\big{)}^{-2} \Big{]}^{1/4}\,{\rm K},
\end{equation}

\noindent 
where $Q_{\rm abs} / Q_{\rm e}$ is the ratio of absorption to emission
efficiency for the grains. For blackbodies (i.e., $Q_{\rm abs} /
Q_{\rm e} = 1$) we would expect a dust temperature around 40\,K,
adopting a stellar luminosity of $5 \times 10^4$\,L$_{\odot}$ and
separation from the star of 1$\farcs$5 or 10,500\,AU (assuming a
distance of 7\,kpc).  Instead, the SED indicates relatively hot dust
at a temperature of $\sim 190$\,K (Figure~\ref{fig:sed}), which would
require $Q_{\rm abs} / Q_{\rm e}$ to be very large, around 500.  This
efficiency could indicate very small grains with radii $<
0.1$\,$\mu$m, which have very low heat capacity and can be superheated
by UV radiation or trapped Ly$\alpha$.  Significant additional heating
might also occur from collisions with the ionised gas, because grains
that are charged due to the photoelectric effect can have a much
larger cross section for collisions with charged particles in ionized
gas. A similar temperature difference occurs in the main equatorial
ring, where the observed dust temperature of 85\,K is much larger than
the expected equilibrium blackbody temperature of $\sim 20$\,K.  If
additional heating from trapped Ly$\alpha$ or collisions with ionised
gas are present, this is important to consider when conducting simple
radiative-transfer models of the circumstellar dust emission around
luminous stars, and adds a note of caution for dust properties
inferred from those models.

In any case, the exact heating mechanism of the dust interior to the
ring is less critical than the fact that it resides there.  The
existence of dust at this location is problematic for conventional
interacting-wind models for the formation of a ring nebula like the
one around SN~1987A (Blondin \& Lundqvist 1993; Martin \& Arnett 1995;
Collins et al.\ 1999; Morris \& Podsiadlowski 2009).  These models
predict that the R-T instabilities in the ring itself mark the contact
discontinuity between the BSG wind and a pre-existing disk wind.  If
so, then the volume interior to the ring should be filled with mass
from the BSG wind.  There will be a reverse shock somewhere between
the star and the ring, but this is not expected to form dust.  Steady
BSG winds do not form dust on their own, and the SED and spectrum of
the central source in SBW1 reveal no evidence for a dusty red giant
star in a binary system.  Thus, the dust in the interior of the ring
around SBW1 must have some other origin, and suggests that a simple
interacting-winds model incorporating only hydrodynamic effects must
be rejected.

\begin{figure*}\begin{center}
    \includegraphics[width=6.0in]{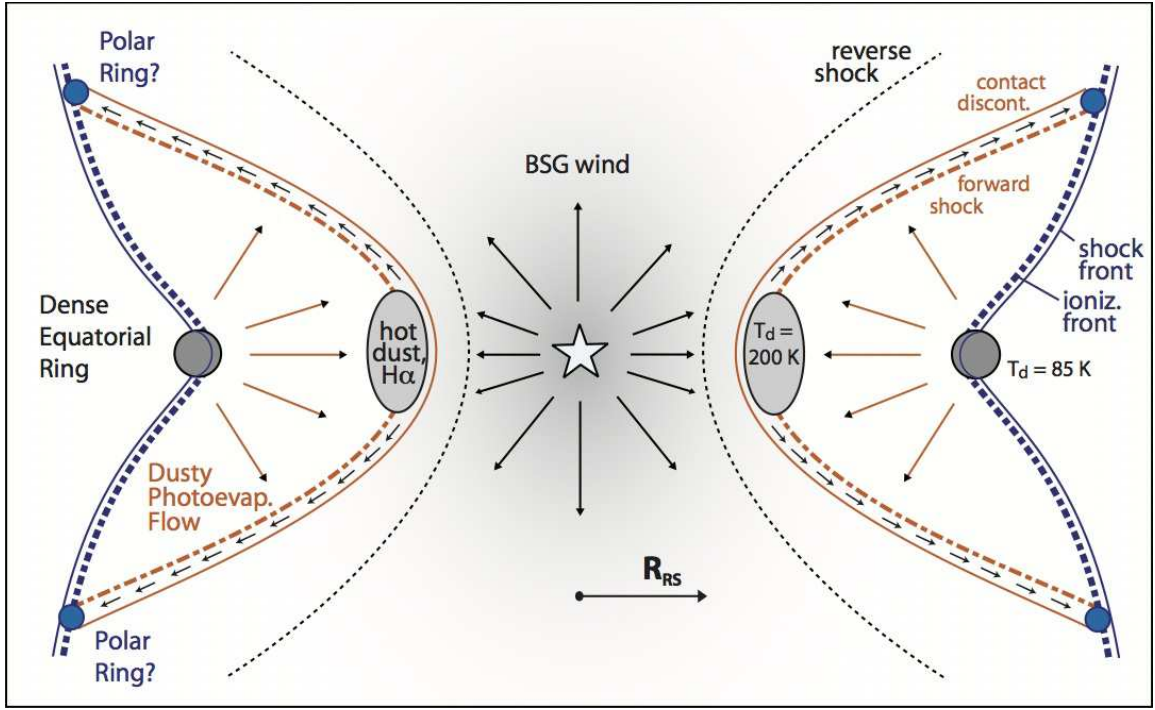}
\end{center}
\caption{A sketch of the proposed emitting geometry of SBW1.  The
  observed features are the dense equatorial ring, the polar rings,
  the outer hourglass-shaped shell (dashed and thin solid blue
  curves), the hot dust and H$\alpha$ that piles up near the contact
  discontinuity in the equatorial plane, and the central star.  Other
  structures are proposed to explain these observed features as
  discussed in the text: the BSG stellar wind, the reverse shock in
  the BSG wind (black dashed), the dusty photoevaporative flow (DPF)
  off the ring (orange arrows), the forward shock driven into the DPF
  (dashed orange), and the contact discontinuity between the shocked
  DPF and the shocked BSG wind (solid orange).  The thin solid blue
  curves represent a weak D-type shock front driven into the neutral
  ring and outer nebula by the pressure of the ionization front, and
  the small black arrows represent the accelerated shear flow of the
  shocked DPF.  Relative sizes of various features are meant to
  demonstrate the general structure, and are not exactly to
  scale. Whether the polar rings actually arise at the intersection of
  the outer hourglass-shaped shell and the shocked DPF is unclear from
  observations; this is a conjecture.  With some adaptation in scaling
  (i.e., opening angle of the reverse shock, latitude of the polar
  rings, etc.), this same geometry might apply to the nebula around
  the progenitor of SN~1987A.}
\label{fig:sketch}
\end{figure*}

Implications for a possible origin of this structure are discussed in
the next section, related to the proposed existence of an ionised
portion of a progenitor RSG wind around SN~1987A (Chevalier
\& Dwarkadas 1995).  In short, we propose that the dusty region
interior to the SBW1 ring originates from photoionisation and
photoevaporation of neutral dusty gas in the dense equatorial ring,
which then expands to partially fill the interior region of the ring.
This dusty photoevaporative flow meets the expanding BSG wind at a
shock interface with a toroidal geometry inside the ring.  A sketch of
the proposed geometry for SBW1 is given in Figure~\ref{fig:sketch},
and is discussed in more detail below.

\section{DISCUSSION}

\subsection{Origin of the Dusty H~II Region Inside the Ring: The Role
  of Ionisation and Photoevaporation}

In the previous sections, we have described how observed structural
properties of the SBW1 ring seem to be in conflict with some aspects
of hydrodynamical models for the formation of SN~1987A's nebula.
Specifically, we highlighted the lack of long R-T fingers and the 
existence of dust interior to the ring.  Both challenge the standard
interacting-winds picture where R-T fingers form as a result of
instabilities in the contact discontinuity between a fast BSG wind and
a flattened disk-like RSG wind.  The presence of dust far interior to
the ring is extremely important, because it means that the BSG wind
(presumably devoid of dust) has not yet reached the ring, and in fact
has not even penetrated past about 1/3 the radius of the ring in the
equatorial plane.

In this section we suggest a modification to this scenario, wherein
ionisation and evaporation of a dense neutral ring produces a
photoevaporative flow which, in turn, profoundly influences the
overall hydrodynamics and shaping of the nebula.  These ionisation
effects have so far been neglected in numerical simulations, but we
argue that they are essential, and our intention is that our
discussion of the observations will inspire such a numerical study.
Our proposed scenario follows the suggestion by Chevalier \& Dwarkadas
(1995), who invoked a dense H~{\sc ii} region arising from
photoionised portions of a previous RSG wind in order to account for
some aspects of the advancing blast wave of SN~1987A.  The additions
and modifications that we make to this scenario concern the dynamical
influence of the ionised photoevaporative flow off the ring, the
nature of the interaction between this flow and the BSG wind, and the
origin of dust in the system (not necessarily from a RSG wind).

In Figure~\ref{fig:sketch} we show a sketch of the proposed geometry
for the SBW1 ring and material inside it (this is a side view in the
equatorial plane).  In brief, the picture we propose is that the warm
dust seen in Gemini South/T-ReCS images resides at or near the
location of a shock front at the interface where the outflowing BSG
wind meets the inward flow of dusty ionised gas that has been
photoevaporated from the dense equatorial ring/torus.  In a two-dimensional 
cross section, this structure has a parabolic shape due to the divergent
flow from the ring.  This is analogous to the opening angle of two
colliding winds in a binary system, except here the structure is
toroidal rather than a cone, because the source of the flow is a dense
ring rather than a companion star.  In effect, this ionised
photoevaporative flow off the dense equatorial ring is able to keep
the BSG wind at bay in the equatorial plane, preventing it from
reaching the equatorial ring so that no direct hydrodynamical
interaction between the ring and the BSG wind is able to occur.
Instead, ionising photons that travel ahead of the shock have the most
important dynamical effect because of the increase in pressure when
the gas is ionised.  The geometry sketched in Figure~\ref{fig:sketch}
could arise from a sequence of events such as that described in the
following section.

\subsection{Formation of the Observed Structure in SBW1}

In this section, we outline a possible sequence of events that may
have led to the formation of the ring nebula around SBW1.  We also
discuss the most relevant physical mechanisms for each component of
the nebula. The key difference between this and previously suggested
scenarios is the hydrodynamic role of an ionised photoevaporative flow
off the ring.  A key point is that the ionised gas pressure is allowed
to have a major impact on the hydrodynamics because of the slow 
(10--20\,km\,s$^{-1}$) initial expansion speed of the dense ring.

\subsubsection{Episodic Ring Ejection as a BSG, Not a RSG Disk-Wind}

Several previous studies that sought to explain the unusual structure
of the nebula around SN~1987A assumed some form of a fast BSG wind
(300--400\,km\,s$^{-1}$) blowing into a slower RSG wind 
(5--10\,km\,s$^{-1}$).  Different versions of this are seen in the more
standard ``hydrodynamics-only'' interacting-winds model (Luo \& McCray
1991; Blondin \& Lundqvist 1993; Martin \& Arnett 1995), as well as in
the H~{\sc ii} region model of Chevalier \& Dwarkadas (1995).  In the
sections to follow, we argue that the structures observed around SBW1
seem to clearly support the model of Chevalier \& Dwarkadas (1995),
wherein the effects of photoionisation are critical in producing a
dense H~{\sc ii} region at low latitudes, but allowing the BSG wind to
expand freely out the poles.  However, one aspect in which our
suggested scenario differs from that of Chevalier \& Dwarkadas (1995)
is in the assumption of a previous RSG wind, as discussed in this
section.

The assumption of a previous RSG wind is motivated by the slow ($\sim
10$\,km\,s$^{-1}$) expansion speed in the equatorial ring of SN~1987A.
However, this assumption also introduces significant difficulties, as
the RSG wind must be concentrated in a thin disk with a very high
equator-to-pole density contrast (Luo \& McCray 1991; Blondin \&
Lundqvist 1993; Martin \& Arnett 1995).  Moreover, these hydrodynamic
simulations predict an hourglass structure with complete polar caps,
rather than empty polar rings.  Since the extended envelope of any
single RSG star will have a very low angular velocity, one must invoke
some interaction with a companion star in a binary system (possibly
even a merger) in any scenario involving a RSG (Collins et al.\ 1999;
Morris \& Podsiadlowski et al.\ 2009).

Yet, if one must invoke a binary system, then a previous RSG phase is
not really needed, because a RSG wind is not the only way to form a
slowly expanding dusty equatorial ring, and moreover, such a flattened
disk-like outflow has never been observed around a RSG (binary or
single).  A different case that has well-established observational and
theoretical precedent is the episodic ejection of a ring by a BSG.
This could have been the result of a binary merger event (Morris \&
Podsiadlowski 2009), an episodic ejection event occurring during 
Roche-lobe overflow (RLOF)
as seen in the eclipsing binary RY~Scuti (Smith et al.\ 2011), or an
eruptive ejection from a rotating star (Smith \& Townsend 2007;
Shacham \& Shaviv 2012).  In either a binary merger event or
accretion from a companion in RLOF, a large amount of angular momentum
must be shed in order for material to be accreted onto the merger
product or the mass-gaining star, respectively, and one thus expects
the mechanical mass shedding to occur in the equator.  If the mass
loss is episodic, this brief equatorial ejection will form a ring
rather than an extended disk because of the brief nature of the mass
ejection.  Shedding of mass at the equator in this way forms a nearly
Keplerian ring with a relatively slow radial expansion speed, which
can be much less than the escape speed from the surface of a BSG star.

A prime example of the scenario we suggest is seen in the Galactic
eclipsing binary system RY~Scuti, in which an OB supergiant binary
system undergoing RLOF has recently ejected a ring/torus that is
spatially resolved in {\it HST} images (Smith et al.\ 1999, 2002,
2011).  The radial expansion speed of the equatorial ring/torus in
RY~Scuti is only 30--40\,km\,s$^{-1}$ (Smith et al.\ 2002), much less
than the several hundred km\,s$^{-1}$ escape velocity or wind speed
from the O9/B0 supergiant stars.  Moreover, this has occurred without
a RSG phase, because the stars are too close together with an orbital
period of only $\sim 11$ days (Grundstrom et al.\ 2007; Smith et al.\
2002, and references therein).  RY~Scuti is an important example,
because measured proper motions of the very young (120\,yr old) nebula
show that the equatorial mass loss was the result of a pair of {\it
  brief episodic ejections} rather than a steady flow of mass from the
equator (Smith et al.\ 2011).

Moreover, the nebula around RY~Scuti
reveals that the slow equatorial outflow led to the formation of
significant amounts of dust (Gehrz et al.\ 2001), even though it is a
binary composed of two hot OB supergiants.  In fact, episodic or
eruptive mass loss like that seen in luminous blue variables (LBVs)
may be an essential ingredient for the formation of dust around BSG
stars, as discussed in detail recently by Kochanek (2011).  The class
of B[e] supergiant stars provides many additional examples of BSG
stars that have formed significant quantities of dust in a slowly
expanding equatorial ring or torus (e.g., Zickgraf et al.\ 1996).
{\it Thus, the formation of a slow and dusty equatorial ring does not
  require a previous RSG phase.}  In fact, of the three known Galactic
analogues of the nebula around SN~1987A (these are HD~168625, Sher~25,
and SBW1), two (Sher~25 and SBW1) have chemical abundances
that are inconsistent with a previous RSG phase (Smartt et al.\ 2002;
Smith et al.\ 2007).  The third analog, HD~168526, is an LBV that has
quite likely suffered previous eruptive mass loss (Smith 2007).

What about the observed clumpy structure of the SBW1 ring?  As noted above,
the dust residing far interior to the ring suggests that the BSG wind
has not yet reached the radius of the dense equatorial ring, and as
such, the clumps observed in the equatorial ring cannot be the result
of R-T instabilities at the contact discontinuity as required in
interacting-winds models.  We suggest instead that this regular clumpy
structure may occur naturally due to thermal instability and
fragmentation of an ejected ring, rather than from R-T instabilities.
In both SBW1 and SN~1987A, the measured ring expansion speed is 
10--20\,km\,s$^{-1}$, so the initially ionised ring would be expected to
fragment as it cools and recombines (and its sound speed drops well
below 10\,km\,s$^{-1}$).  Perhaps this fragmentation could lead to the
fairly regular series of clumps around the ring in both objects
(Figure~\ref{fig:deproject}).  This is, however, a minor point.

Subsequently, as the BSG star now seen at the centre of SBW1 recovers
from the mass-loss event and presumably contracts to become a hotter
BSG, its wind speed (escape speed) and its ionising flux could both
increase, and their ratio has an important influence on the resulting
structure.  The influence of photoionisation of the ring ejecta is
discussed next.

\subsubsection{Photoionisation of the Ring, and a Dusty
  Photoevaporative Flow}

After ejecting a thin equatorial ring, the inner edge of the ring will
be exposed to ionising photons from the central BSG star.  Chevalier
\& Dwarkadas (1995) predicted that photoionisation had a strong impact
on the structure interior to the SN~1987A ring, and we suggest a
similar process in the case of the observed structure around SBW1.

The central star of SBW1 is a B1.5~Iab supergiant, which should produce 
an ionising photon flux of roughly $Q_{\rm H} \approx 10^{48}$\,s$^{-1}$ 
(see Martins et al.\ 2005).  This radiation flux ionises the
inward-facing surface of the ring at a radius $R \approx 0.2$\,pc from
the star.  In \S 3.1, we found that the total mass of ionised gas
emitting H$\alpha$ was too low, inconsistent with the larger mass we
would expect to be associated with the cool 85\,K dust in the ring.  We
therefore conclude that most of the mass in the ring is neutral H
residing in high-density clumps in the ring, which must be
self-shielded from the BSG star's ionising radiation.  The minimum
density $n_H$ required to keep the clumps neutral can be estimated
from simple ionisation balance:

\begin{equation}
%n_H = \sqrt{\frac{Q_{\rm H}}{4 \pi R^2 \alpha_B L}},
n_H = \big{(}{\frac{Q_{\rm H}}{4 \pi R^2 \alpha_B L}}\big{)}^{1/2},
\end{equation}

\noindent 
where $\alpha_B \approx 3 \times 10^{-13}$\,cm$^3$\,s$^{-1}$ is
the Case B hydrogen recombination coefficient and $L = 0.0034$\,pc is
the same observed depth of the emitting layer of the ring as before. We
find an expected density in the ring of $n_e \approx 8500$\,cm$^{-3}$. 
This is more than an order of magnitude higher than the
electron densities derived from the [S~{\sc ii}] line-intensity ratio
and from the observed emission measure of H$\alpha$ (both are
discussed above in \S 3.1), thus confirming that H$\alpha$ and [S~{\sc
  ii}] emission trace only a fraction of the total mass,
associated with a thin ionised skin and ionised photoevaporative flow
coming off the ring (the rest of the mass remains neutral).

When the neutral gas in the clumps gets ionised, its temperature
increases to $\sim 10^4$\,K and the overpressure causes the gas to
expand away from the dense clump (e.g., Oort \& Spitzer 1955).  The
expansion speed should be of the same order as the sound speed in
ionised gas, $c_s \simeq 10$\,km\,s$^{-1}$.  This is also comparable
to the expansion speed of the ring itself.  The photoablation
mass-loss rate is roughly

\begin{equation}
\dot{M} = 2 \pi R L \mu m_H n_e c_s
\end{equation}

\noindent which is roughly equal to 8$\times$10$^{-6}$ $M_{\odot}$
yr$^{-1}$ for the parameters given above.  This, in turn, suggests a
lifetime for the neutral ring of 6$\times$10$^4$ yr, longer than the
$\sim$10$^4$ yr dynamical age of the ring (Smith et al.\ 2007).  The
resulting ionised photoevaporative flow is closely analogous to the
ionised photoevaporative flows from neutral clouds at the edges of
H~{\sc ii} regions, as discussed in detail by Bertoldi (1989) and
Bertoldi \& McKee (1990).  A similar role of photoevaporation is at
work in the proplyds in the Orion Nebula, as discussed more below.

The originally neutral gas in the dense equatorial ring is mixed with
dust.  This cool dust resides outside the ionisation front in the
ring, and is observed as the $\sim 85$\,K component in the IR SED of
SBW1 (Figure~\ref{fig:sed}), which is spatially resolved to be
coincident with the thin ring in our 18\,$\mu$m Gemini/T-ReCS image
(Figure~\ref{fig:trecs}).  After being struck by ionising photons, the
ionised photoevaporated material from the ring will likely entrain the
dust with it as it expands into the interior of the ring.  Thus, we
refer to the resulting expansion as a dusty photoevaporative flow
(DPF) in Figure~\ref{fig:sketch}.  This dust in the DPF moves inward
and eventually piles up at the shock front (discussed in the following
section), giving rise to hot ($\sim 190$\,K) dust emission peaks
inside the ring that are spatially resolved in Gemini images
(Figures~\ref{fig:trecs} and \ref{fig:color}).  This hypothesis seems
far more likely than any alternative explanation for the origin of
dust seen interior to the ring in mid-IR images.  Such alternatives
would require that dust forms quickly in the fast BSG wind and
survives passage through the strong (several $10^2$\,km\,s$^{-1}$)
reverse shock.

This physical situation we propose for SBW1 has an interesting analog
in the evaporating protoplanetary disks (the so-called ``proplyds'')
seen in {\it HST} images of the Orion Nebula (see Bally et al.\ 1998;
Johnstone et al.\ 1998).  In these objects, UV radiation from the
nearby O6~V star $\theta^1$C~Ori causes a photoevaporative flow off
the dense and dusty protoplanetary disk envelopes that are seen in
silhouette.  After passing through an ionisation front, the ionised
photoevaporative flow continues to expand in a divergent flow until it
collides directly with the fast stellar wind of $\theta^1$C~Ori.  The
proplyds located in the Trapezium very close to $\theta^1$C~Ori show
bright arcs of H$\alpha$ emission marking the shock between the
ionised photoevaporative flow and the stellar wind (Bally et al.\
1998).  Interestingly, these same H$\alpha$ arcs are also very bright
in the thermal-IR continuum in high-resolution 11.7\,$\mu$m images
(Smith et al.\ 2005), indicating the presence of hot dust.  Since the
wind of $\theta^1$C~Ori is too fast, rarefied, and hot to form dust on
its own, the dust in these shocks must have been entrained in the
ionised photoevaporative flow from the proplyds.  This is akin to the
situation we propose for SBW1, except that instead of an externally
ionised disk envelope in the Orion proplyds, we have a thin ring
illuminated from the inside.

Because of the thin-ring geometry of the neutral gas reservoir, the
DPF will expand into the interior of the ring toward the star, but the
pressure of the ionised gas will also cause it to expand over a range
of latitudes above and below the equator. This will form a thicker
torus-like geometry stretching from the equatorial plane to mid
latitudes, as depicted by the DPF in Figure~\ref{fig:sketch}.

\begin{figure*}\begin{center}
    \includegraphics[width=4.8in]{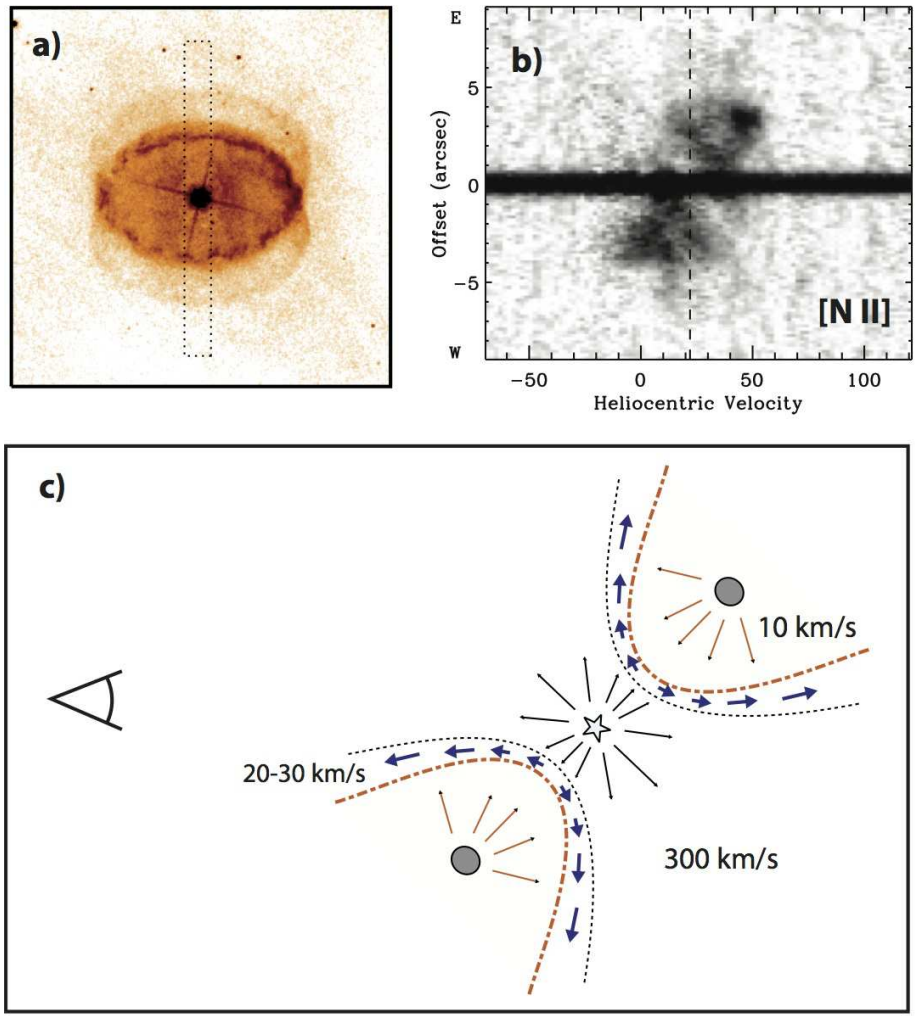}
\end{center}
\caption{A possible kinematic signature of the latitudinal outflow
  structure resulting from the collision between the DPF and the BSG
  wind.  Panel (a) shows the approximate slit position for the
  long-slit echelle spectrum in (b).  Panel (b) is the
  position-velocity diagram for [N~{\sc ii}] $\lambda$6583 from the
  long-slit echelle spectrum of SBW1, as presented by Smith et al.\
  (2007).  (The kinematic emission structure of [N~{\sc ii}] is
  essentially the same as that of H$\alpha$.)  The cartoon in Panel (c) 
  is a simplified version of the sketch in Figure~\ref{fig:sketch},
  concentrating on the shock interface.  Arrows denote the nonradial
  flow of plasma down the shock, in the dense region between the
  forward and reverse shocks.}
\label{fig:emmi}
\end{figure*}

\subsubsection{Collision of the DPF and the BSG Wind}

The DPF will expand away from the dense equatorial ring, primarily
into its interior, until it collides with the outflow from the stellar
wind of the central BSG.  This collision between the BSG wind and the
DPF will have three boundaries: the reverse shock that decelerates the
BSG wind, the forward shock driven into the DPF, and the contact
discontinuity between them. These are labeled in
Figure~\ref{fig:sketch}.  The zone between the forward shock and
contact discontinuity is drawn as a thin region, due to the
high-density gas and slow velocity of the forward shock.  The region
between the contact discontinuity and reverse shock is thicker, due to
the higher speed and lower density of the BSG wind.

Consider the balance between the gas pressure of the ionised
photoevaporative flow (H~{\sc ii} region) and the ram pressure of the
BSG wind in the equatorial plane:

\begin{equation}
n_e k T = \frac{\dot{M} V_{\rm BSG}}{4 \pi R^2},
\end{equation}

\noindent 
where $n_e$ and $T$ are the ionised gas density and
temperature in the H~{\sc ii} region, and $\dot{M}$ and $V_{\rm BSG}$ are
the mass-loss rate and wind speed of the BSG wind, respectively.
While the pressure in the H~{\sc ii} region is roughly constant with
radius, the ram pressure of the wind drops with radius from the star
if we assume a steady BSG wind ($R^{-2}$ density profile). Then $R$ is
the radius where the two balance, given by

\begin{equation}
  R \ = \ 0.05 \ \Big{(} \dot{M}_{-7} \ V_{300} \Big{)}^{1/2} \Big{(}\frac{n_e}{500 \ {\rm cm}^{-3}} \Big{)}^{-1/2} \ {\rm pc},
\end{equation}

\noindent 
where we have assumed $T = 10^4$\,K in the H~{\sc ii} region,
$\dot{M}_{-7}$ is the BSG wind mass-loss rate in units of 
$10^{-7}$\,M$_{\odot}$\,yr$^{-1}$, and $V_{300}$ is the 
wind speed in units of 300\,km\,s$^{-1}$.  
We derived a density of about 500\,cm$^{-3}$ for the
ionised gas in the interior of the ring (\S 3.1), so this is used as a
reference value.  We derived an {\it upper limit} to the BSG mass-loss
rate of $3 \times 10^{-7}$\,M$_{\odot}$\,yr$^{-1}$ from the IR flux of
the central star (\S 2.5), so $10^{-7}$\,M$_{\odot}$\,yr$^{-1}$ is used
as a fiducial value.  We assumed a value for $V_{\rm BSG}$ of 
300\,km\,s$^{-1}$, as above.  These are similar to the values adopted for 
the progenitor of SN~1987A by Chevalier \& Dwarkadas (1995) as well.  With
these values, the stand-off shock will be at $R \approx 0.05$\,pc 
from the BSG.  This is about 25\% of the radius of the
ring, which roughly matches the location of the hot dust and enhanced
H$\alpha$ emission in images, and which we associate with the shock
front.  The agreement indicates that this interpretation is plausible.
An approved program in {\it HST} Cycle 20 will use the Cosmic Origins
Spectrograph (COS) to obtain a UV spectrum of the
central star in SBW1, allowing direct estimates of $\dot{M}$ and
$V_{\rm BSG}$.

Since the H~{\sc ii} region is toroidal and caused by evaporation of a
thin equatorial ring, we expect the density to drop at latitudes above
and below the equator because the ionised photoevaporative flow
diverges.  If so, this would cause the location of the shock front to
curve outward to larger radii, as depicted in Figure~\ref{fig:sketch}.
The variation with latitude is discussed further below.

In the sketch of the geometry shown in Figure~\ref{fig:sketch}, mass
in the outflowing BSG wind and gas from the ionised photoevaporative
flow will both pile up at the shock front.  After colliding, the
post-shock gas should then flow away from the star, in a nonradial
trajectory that follows the geometry of the shock interface (small
black arrows in Figure~\ref{fig:sketch}).  This is analogous to flow
down the shock cone in a colliding-wind binary, but here the geometry
is toroidal rather than a cone.  The material in this post-shock flow
is denser than gas in the H~{\sc ii} region or the BSG wind.  It is
this higher-density zone of ionised gas that probably dominates
H$\alpha$ emission because of the $n_e^2$ dependence of recombination
emission.

In fact, it seems likely that the dense gas in this post-shock flow
has already been observed directly.  There is evidence for the
resulting nonradial flow in high-resolution, long-slit echelle
spectra of H$\alpha$ and [N~{\sc ii}] emission in SBW1 that we
presented in a previous study (Smith et al.\ 2007), obtained with the
EMMI spectrograph on the NTT.  In that paper, we were perplexed by the
unusual position/velocity structures, but it now makes sense in the
context of the geometry proposed above.  In Figure~\ref{fig:emmi} we
show the position-velocity diagram of the earlier EMMI echelle spectra
of [N~{\sc ii}] $\lambda$6583, presented originally by Smith et al.\
(2007), as well as a sketch for how these velocity structures may
arise from the nonradial flow at the shock interface.  While a
quantitative test of this idea will require numerical simulations of
the hydrodynamics, the cartoon of the nonradial flow trajectories in
Figure~\ref{fig:emmi}c provides a reasonable explanation for the
observed position-velocity structure in Figure~\ref{fig:emmi}b.

\subsubsection{Latitudinal Structure}

In both SBW1 and around SN~1987A, {\it HST} images do not reveal dense
nebular material in the polar directions.  Rather, the polar rings of
SN~1987A are true hollow rings residing at mid latitudes (Burrows et
al.\ 1995; Plait et al.\ 1995). The same appears to be true based on
our {\it HST} images of SBW1.  The lack of any dense polar structures
in SBW1 or in SN~1987A argues against a shaping mechanism like the one
discussed by Morris \& Podsiadlowski (2009), because that formation
model predicts dense polar caps in the nebula.  The same is true for
most of the colliding-wind models that predict complete hourglass
shapes (see above), not empty rings at mid latitudes.  If such polar
caps existed, they would be impacted by the fast polar wind of the
BSG, and would likely be swept into a thin dense shell, and this
interface would therefore be easy to detect in deep images.  

Instead, the empty polar regions favour a different scenario like that 
suggested by Chevalier \& Dwarkadas (1995), which is similar to the picture 
we advocate here.  Rather than a pre-existing disk wind, we have proposed
that the central BSG star episodically ejects a dense ring, which is
then photoionised as it expands.  The ionised gas expands to form a
thick torus around the ring, which provides a barrier for the BSG
wind.  The collision between the wind and the ionised torus (H~{\sc
  ii} region) creates a curved shock front.  In this picture, all of the
slow and dense material is confined to low latitudes, so that the BSG
wind is able to expand with no obstruction out the poles.

{\it Is the BSG wind spherical?}  Ejection of an equatorial ring
requires a system with excess angular momentum, suggesting either a
rapidly rotating star or a close binary system.  Even in a binary
system with RLOF, the mass-gainer star accretes both mass and angular
momentum, and would be expected to be in very rapid rotation.  The
geometry of the stellar wind from a rapidly rotating hot star is
complicated, because hot stars have line-driven winds.  Rapidly
rotating stars suffer gravity darkening (von Zeipel 1924), producing
cooler equatorial zones and hotter poles.  Such stars also have higher
escape speeds out the poles than at the equator, so the polar winds
are faster.  The oblate structure of the star produces a larger
cross-section in the polar directions, and also the hotter poles
produce much more radiative flux ($F_{\rm rad} \propto T^4$).  This is
critical for a radiation-driven wind.  This stronger radiative flux at
the poles, combined with velocity-dependent forces, means that rapidly
rotating stars have higher mass flux and higher wind speed out the
poles (see Owocki et al.\ 1994, 1996, 1998; Owocki \& Gayley 1997).
This is for radiatively driven mass loss --- different from the more
intuitive case of mechanical mass loss from a spinning object, which
should produce an equatorial flow (appropriate for RLOF).

Including a stronger polar wind for the present state of the central
BSG would not change the qualitative picture in
Figure~\ref{fig:sketch}, because a stronger wind out the poles would
still escape unobstructed.  It may, however, make it easier to produce
this shape and to reconcile the required values of $\dot{M}$ with
those of normal BSG stars.  Note that for a B1.5~Iab star, the expected 
value of $\dot{M}$ is $\sim 5 \times 10^{-7}$\,M$_{\odot}$\,yr$^{-1}$, 
or perhaps a small factor less if we allow for clumping.
This would be in rough agreement with our upper limit for SBW1 derived
from IR observations, $\sim 3 \times 10^{-7}$\,M$_{\odot}$\,yr$^{-1}$.
However, for both SBW1 and SN~1987A, densities at the equator would
seem to suggest a slightly lower value of $\dot{M}$, closer to
$10^{-7}$\,M$_{\odot}$\,yr$^{-1}$.  Moreover, our assumed wind speed of
300\,km\,s$^{-1}$ (the same value assumed by Chevalier \& Dwarkadas
1995) is somewhat lower than the speeds we expect for a B1.5
supergiant, closer to 500--600\,km\,s$^{-1}$.  Perhaps the factor of 2--3
discrepancy in $\dot{M}$ and $V_W$ is not too concerning, but allowing
the central BSG to have a bipolar wind --- with higher $\dot{M}$ and
$V_W$ out the poles, and lower $\dot{M}$ and $V_W$ at low latitudes
near the equator --- would make the agreement better.

A bipolar wind from the BSG might also help explain another observed
peculiarity of SN~1987A.  Studies of the evolution of light echoes
from SN~1987A have revealed a large bipolar shell, much larger than
the triple-ring system seen in {\it HST} images (Sugerman et al.\
2005; Crotts et al.\ 1995; Wampler et al.\ 1990).  In the geometry
advocated here, there are no polar caps in the inner nebula, so the
strong BSG wind is free to escape out the poles to much larger radii.
Doing so, a bipolar wind could sweep into the surrounding interstellar
medium and create a much larger bipolar shell.

\subsection{Origin of the Triple Rings Around SN~1987A}

We have noted multiple times that while interacting-winds models can
reproduce structures resembling the equatorial rings around SN~1987A
and SBW1 by forming an hourglass shape (Blondin \& Lundqvist 1993;
Martin \& Arnett 1995; Collins et al.\ 1999), the polar rings of
SN~1987A have actually been an enduring problem with no satisfactory
explanation.  The nebula around SBW1 does not exhibit polar rings that
are as prominent as those around SN~1987A, but the detailed geometry
we see in SBW1 may provide some intriguing clues about how the polar
rings might have been formed.

The key difference between previously published interacting-winds
models and the model advocated here is the existence of a dense DPF
off the compact and dense equatorial ring (Figure~\ref{fig:sketch}).
We have argued that its existence in SBW1 is supported by the presence
of dust interior to the thin H$\alpha$ ring (Figure~\ref{fig:color}),
as well as the unusual kinematic structure of the emission lines in
long-slit echelle spectra (Figure~\ref{fig:emmi}).  Chevalier \&
Dwarkadas (1995) argued for the existence of a similar structure
around SN~1987A based on the properties of the advancing blast wave.
This DPF creates a stand-off shock front (reverse shock) in the
equator at $R_{\rm RS} = 0.05$\,pc, but at latitudes above and below
the equator the interface will take a parabolic shape in cross section
(Figure~\ref{fig:sketch}).

How does the existence of this DPF and reverse shock modify the
situation?  In this scenario, one would still expect the formation of
a thin hourglass-shaped nebula as the H~{\sc ii} region tries to
expand into any neutral gas outside it (see Chevalier \& Dwarkadas
1995).  The fast BSG wind will escape in the polar direction and will
sweep away any of this outer nebulosity, {\it but it will be prevented
  from doing so at low latitudes.}  In effect, the DPF and reverse
shock keep the BSG stellar wind at bay, and thereby shield the nebula
at low latitudes.  Therefore, we expect that the thin outer hourglass
structure will remain at mid and low latitudes, as depicted in
Figure~\ref{fig:sketch}.  

Now, recall that the collision between the
DPF and the BSG stellar wind leads to a nonradial flow of dense gas
down the shock front (Figure~\ref{fig:emmi}).  An interesting
consequence of the proposed flow down the shock interface is that the
densest gas will flow outward at a narrow range of latitudes above and
below the equator.  This flow may eventually intersect the hourglass
structure, which in this picture is the outer boundary of the
photoionised cavity -- labeled as the ionisation shock front in
Figure~\ref{fig:sketch}.  Interestingly, the geometrical intersection
of these two curved surfaces in three dimensions is a pair of
plane-parallel rings above and below the equator.  (This would still
be true if the outer ionisation shock front were an hourglass, a
sphere, or a cylinder.)  If there is a density enhancement associated
with this intersection, it may help explain the origin of the polar
rings around SN~1987A and HD~168625.

Alternatively, a similar idea was suggested by Chi\c{t}\v{a} et al.\
(2008) based on their hydrodynamic simulations of a bipolar BSG wind
interacting with a thin spherical shell formed by the terminal shock
of the previous RSG phase.  They assumed that as the star evolves in a
blue loop after the RSG phase, the star must spin up, and that the
wind from the BSG will have a bipolar shape for the reasons noted in
\S 4.2.4.  These authors showed that the collision of a bipolar wind
with a thin shell could lead to a pair of polar rings.  In this
scenario, the existence of the reverse shock could help to stabilise
the latitude range where the bipolar BSG wind and thin shell
intersect, so that the rings would not be as transient as in those
simulations.

We hope that the observations and discussion presented here will
motivate new simulations that include the effects of a DPF.  This
basic picture also allows for some differences among objects, which
may help explain some of the diversity in the structures seen around
SN~1987A, SBW1, Sher~25, and HD~168625.  For SBW1 and SN~1987A, we
expect that the initial structures of the two nebulae were very
similar, but that in SN~1987A, the structure has been modified: the
dense ring has become more fully ionised and the interior dust was
probably vaporised by a UV flash from SN shock breakout.  Even
ignoring the supernova UV flash, however, certain aspects of the
structure can be adjusted.  For example, other central stars may have
different values of $Q_{\rm H}$ or $\dot{M}$, either of which may
change the relative location of the shock between the BSG and the
photoevaporative flow, and may affect the opening angle.  This could
have important implications for understanding when the SN~1987A blast
wave encountered the H~{\sc ii} region (see Chevalier \& Dwarkadas
1995; Gaensler et al.\ 2000).  The stellar luminosity values for SBW1
and Sk$-$69$\arcdeg$202 are very similar, so their mass-loss rates
should be similar, but SBW1 is hotter (B1.5 instead of B3) and has a
higher value of $Q_H$.  This may strengthen the relative importance of
the DPF somewhat in SBW1.  Nevertheless, SBW1 and SN~1987A may trace
one another more closely than in other BSG ring nebulae, such as
Sher~25 and HD~168625, both of which have substantially higher stellar
luminosities.  Nevertheless, one could probably tune the basic picture
presented here to be applicable to those two objects as well.

\section{SUMMARY}

In this paper, we have presented a series of multiwavelength images
and spectra of the nebula around the blue supergiant SBW1, which is
seen in the Carina nebula.  The observations included the first images
and spectra of this target obtained with {\it HST} using WFC3 and
STIS, as well as IR images from {\it Spitzer} and Gemini South/T-ReCS.
Analysis of this multiwavelength dataset has led to several
conclusions about SBW1 and related systems, enumerated below.

(1) {\it HST}/WFC3-UVIS F656N and F658N images of SBW1 reveal a thin,
clumpy ring, but do not show the long R-T fingers that have been
invoked to explain the origin of the hot spots around SN~1987A when
the tips of the R-T fingers are hit by the blast wave.  Instead, the
structure is more akin to beads along a string with little dense
matter in between.  Instead of arising when a wind swept into an
extended thin equatorial disk, as suggested by hydrodynamical modeling
of SN~1987A, the observed structure may suggest an episodic ejection
of a ring that fragmented into clumps as it cooled.

(2) We noted that the episodic ejection of a slow, clumpy ring could
result during a brief phase of RLOF, as seen in the RY~Scuti eclipsing
binary system.  A previous RSG phase is not needed to
explain the slow expansion speed of the ring.

(3) The radius of the ring, size and angular scale of clumps in the
ring, and deviations from the average radius (wiggles) seen in SBW1
are all an excellent match to the structures seen in SN~1987A's ring.

(4) The radial velocity measured in our STIS spectra as well as the
luminosity and extinction indicated by the SED both confirm the large
distance of $\sim 7$\, kpc (Smith et al.\ 2007), meaning that SBW1 is far
behind the Carina Nebula and seen in projection, rather than actually 
being inside it.

(5) High-resolution images reveal complex diffuse emission filling
the interior of the ring.  H$\alpha$ emission seen by {\it HST}
uniformly fills much of the ring interior, with two enhancements in
the H$\alpha$ flux about 1--2{\arcsec} away from the star on both sides
along the major axis.  Much more importantly, mid-IR images obtained
with Gemini South reveal strong dust continuum emission associated
with these same positions.  The warm dust emission residing inside the
ring is critically important, because the dust cannot arise from the
BSG wind.

(6)  The IR SED reveals two distinct dust-temperature components:
$T = 190$\,K dust with a mass of only $10^{-5}$\,M$_{\odot}$, and
$\sim 85$\,K dust with a larger mass of $5 \times 10^{-3}$\,M$_{\odot}$.
This implies a total gas mass for the SBW1 ring of at least 
0.5\,M$_{\odot}$.  High-resolution mid-IR images from Gemini South 
confirm that the cooler 85\,K component arises from dust in the dense
equatorial ring, whereas the warmer 190\,K dust comes from the
double-peaked diffuse emission in the ring's interior.

(7) Both H$\alpha$ and mid-IR images reveal a clear deficit of diffuse
emission within 1--2{\arcsec} of the central star, suggesting that a
cavity has been cleared by the BSG wind.

(8) We propose a model for the origin of the observed structures in
SBW1 that departs markedly from the standard interacting-winds model
for SN~1987A.  Instead of the equatorial ring arising as a consequence
of a fast wind sweeping into a slow extended disk wind, we propose
that the ring was ejected in an episodic event and then photoionised
by the BSG.  When the dense neutral clumps in the ring get ionised,
they produce an ionised photoevaporative flow that has an important
hydrodynamic effect because of the pressure of the ionised gas.  The
double-peaked diffuse H$\alpha$ and mid-IR emission inside the ring
marks the location of a shock where the BSG wind collides with the
dusty photoevaporative flow (DPF).  This DPF structure is analogous to
the ``H~{\sc ii} region'' proposed by Chevalier \& Dwarkadas (1995) to
account for the expansion of the blast wave of SN~1987A.

(9) In this model, the DPF keeps the BSG wind at bay in the equatorial
plane, thereby shielding the ring and other nebular structures at low
latitudes.  The BSG can expand out the poles unobstructed.

(10) When the DPF and BSG wind collide in the equator, they must flow
downstream.  The shock front probably takes on a curved shape,
diverting the flow to a narrow range of mid-latitudes (see
Figures~\ref{fig:sketch} and \ref{fig:emmi}).

(11) We note some implications for the origin of structures around
SN~1987A.  If the BSG wind expands freely out the poles (i.e., there
are no polar caps in the nebula), it may help explain the existence of
a much larger bipolar shell seen with light echoes (Sugerman et al.\
2005).  Moreover, the nonradial post-shock flow that is confined to a
narrow range of latitudes (point 10 above) may help explain the
existence of the polar rings around SN~1987A when this flow intersects
the dense shell at the outer boundary of the H~{\sc ii} region.  We
encourage further investigation of this idea using numerical
simulations that account for the effects of a photoionised flow.  As
we note in the text, this photoionised flow is permitted to have an
important dynamical effect because the initial expansion speed of the
neutral dense ring is slow, comparable to the sound speed in the
ionised gas.

\smallskip\smallskip\smallskip\smallskip
\noindent {\bf ACKNOWLEDGMENTS}
\smallskip
\footnotesize

Support was provided by the National Aeronautics and Space
Administration (NASA) through {\it HST} grants GO-11637, GO-11977, and
AR-12623 from the Space Telescope Science Institute, which is operated
by AURA, Inc., under NASA contract NAS5-26555. Additional support for
this work was provided by NASA through awards issued by JPL/Caltech as
part of {\it Spitzer} programs GO-3420, GO-20452, and GO-30848.
A.V.F.\ is also grateful for financial assistance from NSF grant
AST-1211916.  This publication makes use of data products from the Two
Micron All-Sky Survey, which is a joint project of the University of
Massachusetts and the Infrared Processing and Analysis
Center/California Institute of Technology, funded by NASA and the
National Science Foundation (NSF). We have also used data products
from the Wide-field Infrared Survey Explorer, which is a joint project
of the University of California, Los Angeles and the Jet Propulsion
Laboratory/California Institute of Technology, funded by NASA.

\end{document}